%% file: MGPaper.tex
\documentclass[10pt,journal,compsoc]{IEEEtran}
\usepackage[most]{tcolorbox}

\ifCLASSOPTIONcompsoc
  \usepackage[nocompress]{cite}
\else
  \usepackage{cite}
\fi
\ifCLASSINFOpdf
\else
\fi
\usepackage{url}

\usepackage{fontawesome}
\usepackage{todonotes}
\usepackage{boldline}
\usepackage{wrapfig}

\usepackage{lineno}

\begin{document}
%


\title{Dealing with Data Challenges when Delivering Data-Intensive Software Solutions}
%
%
%
%

\author{Ulrike M. Graetsch,
        Hourieh Khalajzadeh,
        Mojtaba Shahin,
        Rashina Hoda,
        and John Grundy
\IEEEcompsocitemizethanks{\IEEEcompsocthanksitem U M Graetsch, R Hoda and J Grundy are with the Faculty of Information Technology, Monash University, Melbourne, Australia.  
\protect\\
E-mail: ulrike.graetsch@monash.edu, rashina.hoda@monash.edu, and john.grundy@monash.edu
\IEEEcompsocthanksitem H. Khalajadeh is at Deakin University Email: hkhalajzadeh@deakin.edu.au
\IEEEcompsocthanksitem M. Shahin is with the School of Computing Technologies, RMIT University, Melbourne, Australia, Email: mojtaba.shahin@rmit.edu.au .}
\thanks{Revised Manuscript submitted March 2023}}

%
%

%
\markboth{Submitted to IEEE Transactions on Software Engineering}%
{Shell \MakeLowercase{\textit{et al.}}: Bare Demo of IEEEtran.cls for Computer Society Journals}



\IEEEtitleabstractindextext{
\begin{abstract}
 The predicted increase in demand for data-intensive solution development is driving the need for software, data, and domain experts to effectively collaborate in multi-disciplinary data-intensive software teams (MDSTs). We conducted a socio-technical grounded theory study through interviews with 24 practitioners in MDSTs to better understand the challenges these teams face when delivering data-intensive software solutions. The interviews provided perspectives across different types of roles including domain, data and software experts, and covered different organisational levels from team members, team managers to executive leaders. We found that the key concern for these teams is dealing with data-related challenges. In this paper, we present the theory of dealing with data challenges \textcolor{black}{that} explains the \textit{challenges} faced by MDSTs including gaining access to data, aligning data, understanding data, and resolving data quality issues; the \textit{context} in and \textit{condition} under which these challenges occur, the \textit{causes} that lead to the challenges, and the related \textit{consequences} such as having to conduct remediation activities, inability to achieve expected outcomes and lack of trust in the delivered solutions. We also identified \textit{contingencies} or strategies applied to address the challenges including high-level strategic approaches such as implementing data governance, implementing new tools and techniques such as data quality visualisation and monitoring tools, as well as building stronger teams by focusing on people dynamics, communication skill development and cross-skilling. Our findings have direct implications for practitioners and researchers to better understand the landscape of data challenges and how to deal with them. 
\end{abstract}

}

\maketitle


%
\IEEEpeerreviewmaketitle


%
%
%
%

 
\IEEEraisesectionheading{\section{Introduction}\label{sec:introduction}} 
\IEEEPARstart{S}{oftware} solutions that combine large-scale data analytical functionality and business applications are becoming pervasive\cite{Gartner2021}. Delivery of such solutions involves textcolor{black}{expertise and} skills from \textcolor{black}{different} disciplines including domain expertise, software engineering, data science and cloud computing. \textcolor{black}{These teams are characterised as multi-disciplinary teams because they have different bodies of knowledge, research communities, ways of working, education and career pathways. In a multi-disciplinary team, these differences are not integrated and team members 'function as independent specialists'\cite{med2006multidisciplinarity}.}  \textcolor{black}{In this paper, we refer to these large-scale data analytics applications as Data-Intensive (DI) solutions and use this term broadly to mean systems that analyse and manipulate data to provide predictions and insights. DI systems rely on data, not just algorithms or programs to deliver an outcome or result. 
 Examples of data-intensive systems include imaging diagnostic systems, real estate price predictors, vehicle telemetry systems and business planning software.}
    
How teams building such data-intensive systems with these multi-disciplinary skills are working together has been studied in leading technology organisations, such as Microsoft\cite{amershi2019software, kim2017data} and IBM\cite{Piorkowski2021, Muller2019}. These studies offer insights and recommendations but they may not be similarly feasible or relevant to other organisations. Research studies beyond such big tech organisations have identified challenges regarding multi-disciplinary collaboration, data-quality, communication and adoption challenges \cite{Figalist2020, Khalajzadeh2020, Nahar2021CollaborationCI, aho2020}. \textcolor{black}{To the best of our knowledge, there is a lack of empirical studies that explore the phenomena of multi-disciplinary data-intensive software teams (MDSTs), articulates their challenges and explores the wider context in which these challenges are experienced. Contemporary practitioner perspectives on} key challenges can provide valuable insights and directions for contextualising, prioritising and shaping future work in this area.

To this end, we designed an exploratory study to \textcolor{black}{gain a better understanding of the challenges that MDSTs outside the big tech organisations face.} We sought to explore a diverse range of organisations and perspectives of technical and non-technical team members. We used the Socio-Technical Grounded Theory (STGT) method \cite{Hoda2021}, to collect and analyse data from 24 semi-structured interviews with current industry practitioners in different technical and non-technical roles working with and in MDSTs. Interview participants worked in Europe, Australia, New Zealand, and China across different types of government and non-government organisations.\par
 
We found \textcolor{black}{a high prevalence of data challenges that MDSTs need to deal with to deliver DI solutions}. These include  gaining access to data, understanding data, aligning data, and resolving data quality challenges. Dealing with data challenges often has negative impacts on solution delivery. \textcolor{black}{Focusing on MDSTs that need to deal with data challenges,} we identified relevant contextual factors in MDSTs including the role of data-intensive solutions in the organisation, team member personalities and disciplines. 
We also identified causes \textcolor{black}{that are attributed to MDSTs having to deal with data challenges} including poor data analytics literacy, and lack of adequate description of data, and issues with inter-team dependencies. We investigated how MDSTs dealt with data challenges. At the long term, strategic level, leaders are implementing data governance, and aligning data strategies with business plans. Software engineers are designing new solutions with data challenges in mind and are implementing data quality analysis and monitoring tools. Data experts are using tools to monitor and visualise data quality, and are prioritising and curating data sets. Team leaders are making efforts to improve communication and explanation skills, and are encouraging cross-skilling and positive team cultures to develop high performance teams. \par
 
This paper makes the following contributions:
\textcolor{black}{(a) a description of the causes that contribute to MDSTs having to deal with data challenges (b) the adverse consequences faced by MDSTs, (c) Strategies, tools and skills used by MDSTs to address causes and challenges, (d) description of our application of STGT method, including the 6Cs theoretical model to visualise the theory and (e) a discussion of our results in the context of literature.}

The paper is organised as follows:
Section \ref{sec:related work} provides an overview of related studies and Section \ref{sec:researchmethod} presents our research methodology. In 
Section \ref{sec:findings}, we present our study findings. Section \ref{sec:discussion} discusses our findings in the context of the related literature, identifies key implications and recommendations for practitioners and researchers, and future work. This is followed by presenting an evaluation of our study and the associated threats and limitations in Section \ref{sec:threats} and Section \ref{sec:conclusion} provides a conclusion of the paper.

\section{Related Work}
\label{sec:related work}
The skills of data scientists\cite{Harris2013} and the nature of data science work processes, tools used and collaboration challenges have been explored by studies within several big-tech organisations, such as IBM and Microsoft\cite{Piorkowski2021, Zhang2020, Muller2019, kim2017data}. Piorkowski et al. conducted a case study on communication challenges faced by AI software developers in multidisciplinary teams.
They identified challenges including knowledge gaps between AI/ML and domain experts, having to build trust across disciplines and managing stakeholder expectations \cite{Piorkowski2021}. Muller et al. conducted a Grounded Theory interview study with 21 IBM data science professionals to explore the relationship between data scientists and the approaches and interventions they performed on data -- ranging from accepting the data as given through to actively shaping and creating data \cite{Muller2019}. 

Teams at Microsoft have also been explored to assess how the role of data scientists fits into the traditional way of delivering software and identified the challenges that software engineers, data scientists, and data analysts face when working in multi-disciplinary teams \cite{amershi2019software, kim2017data}. A survey of data scientists and software engineers at Microsoft about the maturity of their delivery approach analysed challenges with building large scale Machine Learning applications and documented key lessons learnt \cite{amershi2019software}. Kim et al. surveyed data scientists at Microsoft and identified ``people related'' issues, including ``explaining the value of data science to software engineers'', ``getting buy-in from engineering teams to collect high quality data'' and, ``communicating insights to leaders effectively'' as challenges to be addressed \cite{kim2017data}.

Beyond these big-tech company based studies, there are also studies covering varying sized and domain organisations. Aho et. al.\cite{aho2020} conducted a small 6 interviews study with data scientists from different consulting organisations  about the nature of data science projects and alignment with software development processes. They identified the multi-disciplinary team as a key characteristic of a data science project, but only covered data scientist perspectives and did not explore the challenges in detail. Multiple team member perspectives were explored by Figalist et. al.\cite{Figalist2020} used a mixed-methods study about why AI is not being used beyond prototyping in Software Analytics projects. They identified data quality and communication challenges as the key cause for not being able to demonstrate value and they describe a `vicious circle' of five key factors including a) lack of time and resources to address data analytics solution delivery challenges, b) low data quality c) inability to bridge cross cultural gaps between team members d) inability to deliver effective prototype analysis and e) inability to provide value \cite{Figalist2020}. Khalajzadeh et al. \textcolor{black}{discussed three industrial case studies that impact the development of data analytics software solutions}, including a) non-technical team members do not have data science or software engineering background, b) technical team members lack domain knowledge, c) data scientists lack software engineering knowledge, d) lack of common language between team members, e) evolution of solution after deployment is poorly supported and f) lack of reuse \cite{khalajzadeh2021bidaml}. Nahar et al. reported an in-depth study of machine learning enabled software systems and interviewed 45 practitioners from multiple disciplines and identified collaboration challenges regarding requirements and planning, training data, and product-model across a range of organisation types\cite{Nahar2021CollaborationCI}. 

\textcolor{black}{Two studies in this related work used grounded theory methods~\cite{Muller2019},\cite{Nahar2021CollaborationCI}. They focused on the process of dealing with data and collaboration challenges in machine learning teams, respectively. Both also conducted extensive literature reviews to ascertain theoretical foundations to pre-inform and provide a lens for their studies. Our exploratory study, on the other hand, applies a socio-technical lens to study how MDSTs deal with data challenges, emerging from the primary data we collected.} \textcolor{black}{We developed an exploratory, Socio-Technical Grounded Theory method based (STGT) study to allow us to start with an open mind, explore the phenomena of MDSTs outside big tech organisations, construct the concepts and categories from the data and guide the selection of the key category for holistic analysis.}\par  

\begin{figure*} [h]
\includegraphics[width=1.0\textwidth]{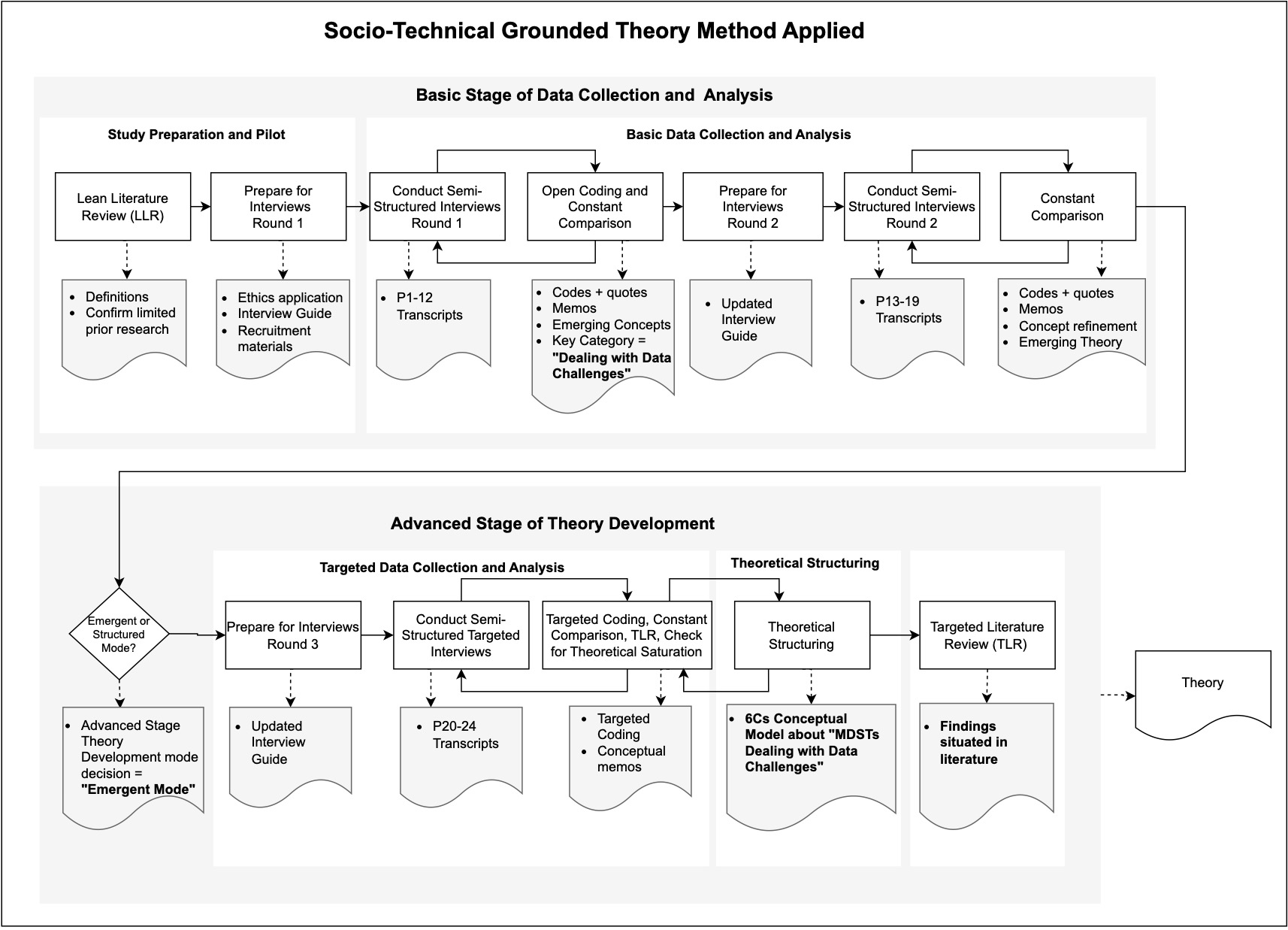}
\textcolor{black}{\caption{Socio-Technical Grounded Theory method applied including Basic Stage Data Collection and Analysis and with "Emergent Mode" applied during Advanced Stage Theory Development.}}
\label{fig:STGTmethod}
\end{figure*}
\input{tables/table1}

\section{Methodology}
\label{sec:researchmethod}
Socio-Technical Grounded Theory (STGT) Method \cite{Hoda2021} \textcolor{black}{uses a primarily inductive approach to generate theory from data. Grounded theory methods, including STGT start with a broad area of interest or topic and progressively narrows down to a key phenomenon or categories that represent the most significant concern of the participants. Using STGT method}, our study commenced with a high level objective to explore the challenges that multi-disciplinary teams face when they deliver data-intensive solutions. \textcolor{black}{Through the analysis of the data collected in the early parts of the study, it quickly became apparent that the most significant challenge faced by the participants was dealing with data challenges. Applying theoretical sampling, we targeted our data collection and analysis on this key category and continued to develop it comprehensively in the advanced stages of theory development. Thus, the overall research question answered in this study is captured as, \textit{What data challenges do multi-disciplinary teams face when delivering data-intensive solutions and how do they deal with them?} Through the analysis and structuring, we were able to identify and conceptualise a number of concepts related to the key category in terms of the \textit{context} in which it occurs, the \textit{conditions}, the \textit{causes} leading up to it, the adverse \textit{consequences}, and the \textit{strategies} applied to address them, giving rise to a grounded theory of how MDSTs deal with data challenges.}


\subsection{Why Socio-Technical Grounded Theory Method?} 
While traditional Grounded Theory methods ~\cite{GlaserBarney2017Tdog,CorbinJulietM.2008Boqr,CharmazKathy2014Cgt} are established research methods in the sociology domain and have been applied in socio-technical domains, STGT has been developed by Hoda\cite{Hoda2021} over a number of years and has recently been formalised and published as a specialised method for technology-intensive domain studies such as software engineering. STGT method addresses the often contentious differences between the traditional Grounded Theory approaches and reconciles and extends them with guidelines for the different types of studies more relevant to software engineering. 

\textcolor{black}{STGT method has been selected for this research study because our research parameters closely align with the four dimensions of the socio-technical grounded theory research framework, as listed below:}\par 

\begin{itemize}
    \item \textcolor{black}{\textit{Socio-technical phenomenon:} Our study's exploration of data challenges facing multi-disciplinary teams and how they deal with them is a socio-technical phenomenon because the experiences of team members are shaped by interactions with the other people in teams and influenced by the technical systems, processes and organisational structures within which they work.}

    \item \textcolor{black}{\textit{Socio-technical domain and actors:} The domains of data-intensive software systems are socio-technical domains ``\textit{with tight coupling between its social and technical aspects}'' \cite{Hoda2021}. Similarly, the actors in these domains, the members of the MDSTs comprising of software engineers, data scientists, AI/ML experts, and domain experts, embody social and technical skills, interactions, and experiences. These team members are not only users of technology, they have a large role in shaping the area that is being explored.}

    \item  \textcolor{black}{\textit{Socio-technical researchers:} The combined experience and skill sets of the research team cover the requisite domain knowledge, philosophical foundations, and qualitative research skills. The interviews were conducted by an experienced industry practitioner/early researcher and supported by a supervisory team with strong data analytics and research skills.  Socio-technical researchers have technical and domain knowledge that enable them to interact with and understand the language and processes that participants refer to.}

    \item  \textcolor{black}{\textit{Socio-technical data tools and techniques:} The research utilised a number of tools including Otter.ai's automated transcriptions,  QSR's NVivo tool to assist with the coding and analysis of interviews, and Zoom recordings.}

\end{itemize}

\textcolor{black}{At all stages of a STGT research study} the research paradigm is a very important consideration because it \textcolor{black}{impacts the approach to interviewing, concept development} and approach to theory development. Research paradigms such as positivism, constructivism and symbolic interactionism have been applied in Grounded Theories \cite{Hoda2021}. The interviewer in this study has extensive industry experience and in data intensive software development and as such the adoption of a subjective, context specific, constructivist research paradigm needs to be acknowledged.  As an interviewer, the researcher had a key role in constructing questions, interpretation of answers, making subjective decisions on interview navigation, and then formulation of concepts and categories and their interrelationships.

\subsection{Applying STGT}
STGT provides a two-step approach to analysing the phenomenon into two stages - the \textit{basic stage} for data collection and analysis and the \textit{advanced stage} for theory development. The separation provides logical points at which findings can be shared \textcolor{black}{and} communicated and \textcolor{black}{the study can either} continue through to theory development, \textcolor{black}{or end}. This study completed both steps and presents a \textcolor{black}{theory about MDSTs dealing with data challenges}. See Figure \ref{fig:STGTmethod} for a schematic overview of the STGT method procedures and steps applied in study.

\subsubsection{Basic Stage of Data Collection and Analysis}

\label{sec:background}
The basic stage of STGT calls for a \textit{lean literature review}, to identify basic research gaps and search for relevant definitions\textcolor{black}{\cite{Hoda2021}. To this end,} we conducted an \textcolor{black}{informal lightweight} literature review, \textcolor{black}{presented} in the Introduction and Related Work sections to summarise related studies in the area of MDSTs that explored issues from a wide range of perspectives.
\textcolor{black}{To prepare for interview based data collection, the first author prepared an initial interview guide to guide semi-structured interviews. The interview questions for the Basic Stage Data Collection and Analysis can be found at\cite{interviewquestions}}. 
The study was approved by the \textcolor{black}{Monash University} Ethics Committee. Prior to commencing interviews with industry practitioners, the first author conducted one \textcolor{black}{pilot} interview with an experienced researcher who had relevant industry experience. As a result of the \textcolor{black}{pilot} interview the first author made minor adjustments to the order and structure of the questions. The \textcolor{black}{pilot} interview was not included in the analysis.\\

\begin{figure*}[h]
\includegraphics[scale = 0.17]{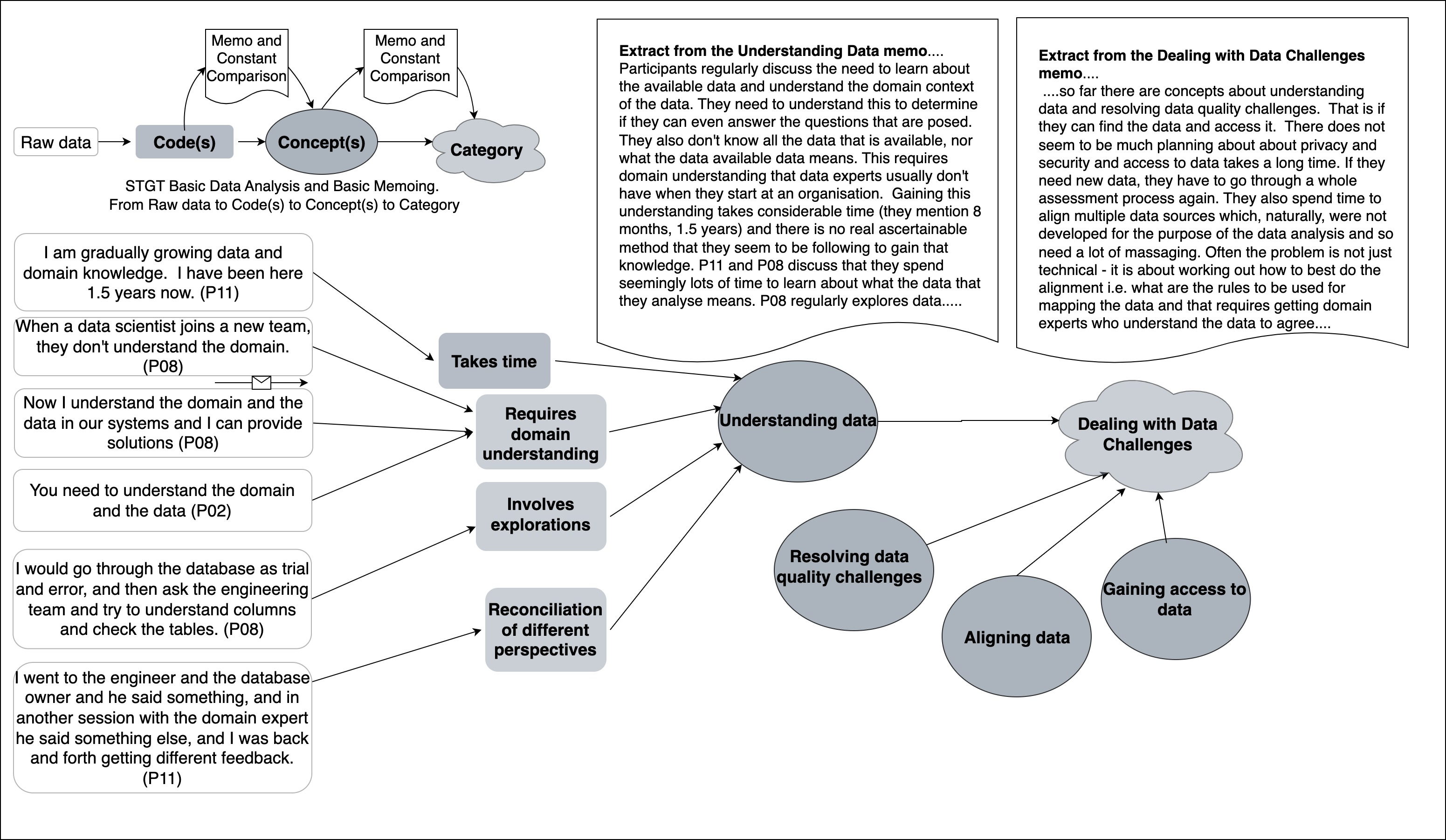}
\textcolor{black}{\caption{Example of the application of STGT Basic Stage Data Analysis to develop the Understanding Data Concept and Dealing with Data Challenges Category from coded raw data using basic memos."}
\label{fig:STGTcode}}
\end{figure*}

\noindent\textcolor{black}{\textit{3.2.1.1 Sampling and Recruitment}}\\
\textcolor{black}{\textcolor{black}{The STGT method supports different types of theoretical sampling processes to target specific data sources in response to identified codes to collect data to identify new concepts, help fill theoretical gaps and saturate concepts. Supported theoretical sampling processes include purposive, random and convenience \cite{Hoda2021}. In our study, we utilised convenience sampling to begin with, followed by \textit{theoretical sampling} as findings started to emerge}. The initial participants were recruited from the industry network of authors. We purposely sought out participants with different roles and levels and conveniently used authors' networks and referrals to approach potential participants. This was combined with a quasi-random approach through advertisements on social media.}\\ 

Interview participant \textcolor{black}{recruitment was interleaved with analysis} throughout the study. \textcolor{black}{Participants were recruited on a voluntary basis and not reimbursed for their time. Overall, the social media advertisements on Twitter and LinkedIn resulted in 1 direct recruit, and 6 referrals, out of which 5 participated i.e. a total of 6 participants. The first author also sent 30 targeted messages using LinkedIn where the first author reviewed her LinkedIn contacts’ job titles and selected contacts that based on their job titles worked in data-intensive job roles. She brought the advertisement posting to the attention of the contact with a personalised, private message and highlighted the nature of the research and suggested that they may be able to provide valuable insights into the study or provide contact details for people that may be able to participate. As a result of these messages and follow up exchanges about their potential suitability or referral to other contacts, 12 participants were confirmed. Of these participants, 10 were direct contacts of the first author and 2 were referrals. The second and fifth authors provided email an additional 9 contact details with relevant job titles who worked in data-intensive job roles. The first author sent recruitment materials and a personalised message to each of the 9 contact points to highlight the nature of the study and the types of roles that we would be looking for to seek contributions. 6 of the 9 contacts progressed to become participants in our study.}\par

In total, 24 participants were recruited and interviewed over a period of 12 months.  Table \ref{tab:participants} outlines basic information about each interview participant including their age, gender, current role, the organisation domain, whether the organisation's core business is development of software services or products, their highest academic qualification, and the number of years and type of professional experience they have. Due to the commercial sensitivity of the information provided and requirement to protect participants identities, details about their organisation and some role titles have been generalised to preserve anonymity.


\textcolor{black}{The potential of a prospective participant to contribute to the study was assessed during recruitment, prior to scheduling an interview. During recruitment, prospective participants were informed that the interviews would be discussing their recent experiences in data-intensive solution delivery and their experience working in multi-disciplinary teams. Not all prospective participants who initially expressed interest during recruitment proceeded with interviews. Some provided explanations including that their experience was not delivery related (for example they were now mainly involved in sales), their experience was not sufficiently recent (for example they had not been involved in delivery over the last 12 months) or they did not work in a suitable team environment (for example they were working as a sole practitioner developing analytics solutions).  Whilst the recruitment had no geographic restrictions, most participants were located in Australia and New Zealand.}\\

\noindent\textcolor{black}{\textit{3.2.1.3 Basic Data Collection and Analysis }}\\
\textcolor{black}{During the basic stage, the first author conducted semi-structured interviews and followed the basic data collection and analysis process over two rounds.}
Due to the COVID-19 pandemic restrictions, all interviews were conducted over Zoom. Interviews during the Basic DCA stage were scheduled to go for between 45-60 minutes, and actually ranged from 40 minutes to more than 2 hours, depending on the topics and depth of discussion.

Open Coding and analysis were completed in Nvivo ~\cite{Hutchison2010, Soliman2004}.  
Round one interviews P1-12 were open coded by reviewing each interview sentence by sentence, comparing to previous codes, and linking interview quotes. \textcolor{black}{When open coding, the first author sought to understand what was happening and what actions were taking place in the account that was provided by the participant\cite{CharmazKathy2014Cgt}}. During coding of each interview, an ''interview memo'' was created to capture thoughts and key elements of the interview.  Additional \textcolor{black}{''basic} memos'' were created to capture items of interest across multiple interviews, relating the codes to each other into concepts and relationships and comparing and contrasting experiences. 

\textcolor{black}{An example of open codes assigned and basic memos is shown in Figure \ref{fig:STGTcode}}.  This information was represented using links within NVivo. Where appropriate, codes were merged. NVivo diagrams were used to visualise mapping of concepts and their linked codes.  The first author regularly met with the others to discuss the approach, review coding examples and progress. The first author regularly exported the code book and memos (which contained the linked codes and quotes) and used the conceptual modelling tool in NVivo to visually model the concepts and categories, providing a visualisation that integrated memos, codes, concepts and categories to the supervising researchers and all the researchers reviewed these collaboratively.

\textcolor{black}{During the initial interviews, a number of challenges were raised by the participants including concepts relating to having to deal with data challenges, delivery challenges relating to productionising and maintaining DI systems as well as project team challenges such as changes of team members. Applying theoretical sampling, the data collection and analysis was progressively narrowed to focus the key participant concern that was emerging from the inductive analysis process, i.e. dealing with data challenges. This included revising the interview questions to focus on the emerging concepts and relationships, instead of remaining completely open as before. Our memos had started to become more conceptual rather than descriptive. At this stage, with the help of the fourth author, an experienced grounded theorist, the team had confidence that a number of concepts surrounding our key category were quite detailed and no new insights were emerging and we decided to proceed to the advanced stage of theory development.}

\subsubsection{Advanced Stage of Theory Development}
The Advanced Stage of \textcolor{black}{Theory Development} provides two options for theory development: the emergent and the structured modes.  \textcolor{black}{Both approaches result in a mature theory outcome. The differences between the two modes are detailed in the guidelines paper \cite{Hoda2021}. While closing out the Basic Stage Data Collection and Analysis, the team assessed that the formation of concepts and categories was quite strong, however, a clear theory structure was as yet not apparent. This offered an opportunity to progress and further develop relationships using Advanced Stage of Theory Development with the 'emergent mode'.}\\


\noindent\textcolor{black}{\textit{3.2.2.1 Targeted Data Collection and Analysis}}\\
\textcolor{black}{Using the emergent mode, we applied \textit{targeted data collection and analysis}, targeting further data collection and analysis toward deepening the already emerging concepts and relationships and deepening our understanding of solution concepts and their relationships to the key category. For example, the final five interviews focused on collecting data about existing concepts, targeted coding at existing concepts and added further details to relationships. Interviews in the advanced stage} were around 20 minutes, reflecting the more targeted nature of the conversations. \\

\noindent\textcolor{black}{\textit{3.2.2.2 Theoretical Saturation and Structuring}}\\
\textcolor{black}{The guidelines stipulate that when data collection ``\textit{does not generate new or significantly add to existing concepts, categories or insights, the study has reached theoretical saturation}'' \cite{Hoda2021}. We found that interviews P23 and P24 strengthened or elaborated concepts and provided good examples, but they did not add new findings to concepts or relationships, thus leading to the conclusion that we had reached theoretical saturation.}

\textcolor{black}{At this stage, based on the concepts derived, the fourth author suggested that \emph{Glaser's 6Cs}\cite{Glaser1978TS} theory template may offer a good fit to support theoretical structuring of the identified concepts in this study.} The 6Cs model presents a key category in terms of six categories: Contexts, Conditions, Causes, Covariances, Consequences, and Contingencies\cite{Glaser1978TS} and has been used previously in SE research (e.g. \cite{hoda2011impact, jovanovic2017transition, masood2020emse}).  \textcolor{black}{In the 6Cs model, the key category is placed at the centre of the model and is then explained in terms of its 6Cs i.e. its Causes, Consequences, Contingencies, Context, Condition and Co-variances. The Causes can represent the reasons for the key category as identified in the data, but the causes is also a mechanism to temporally order data and reflect ''one thing leading to another'' \cite{Glaser1978TS}. Consequences arise out of the key category or come after the key category has been experienced. Contingencies represent strategies and solutions that are being implemented to address the key category. Conditions represent the qualifiers under which the key category is experienced. Co-variances facilitate the inclusion of relationships but not attributing a causal or temporal relationship. Context is described as synonymous with ambience \cite{Glaser1978TS}, i.e. the environment in which the phenomenon occurs.} 

We structured the concepts and relationships into the 6Cs coding model to visualise the theory and found that the data fit the 6Cs model well. \textcolor{black}{For example, all the concepts were identified as leading to the data challenges were grouped under the “causes”. Similarly, all concepts that represented strategies to deal with the data challenges were grouped under “contingencies”, and so on.}\\


\noindent\textcolor{black}{\textit{3.2.2.3 Targeted Literature Review (TLR)}}


\textcolor{black}{Once the findings were solidified and it was time to write them up, we felt the need to situate our findings in light of existing related works. Since the findings of an STGT study emerge inductively from evidence, it is mainly possible to look for highly relevant related works in the advanced stages of the study, and to compare them to the emergent concepts and category. In an STGT study, this need is met with a step called the \textit{targeted literature review}. A TLR is defined as an ``\textit{in-depth review of literature targeting relevance to the emerging/emergent categories and hypotheses, performed periodically during the advanced stage}'' \cite{Hoda2021}. It is typically performed as an informal review, but can also be more systematic, if desired. It is not intended to be a standalone and comprehensive study of literature such as a systematic literature review (SLR) may achieve \cite{kitchenham2002}, although researchers may follow up their STGT study with an SLR where appropriate. In our case, once the main findings were more or less established, a targeted literature review was conducted using using keywords from our emerging concepts (e.g. Data Quality, Data Governance, Cross-skilling).  Closely related works identified through this process were compared with the findings.  The results of this process make up parts of the Discussion section \ref{sec:discussion}.}


\textcolor{black}{After completing Advanced Stage Theory Development we now present our findings.} 

\begin{figure*}[h]
\includegraphics[width=1.0\textwidth]{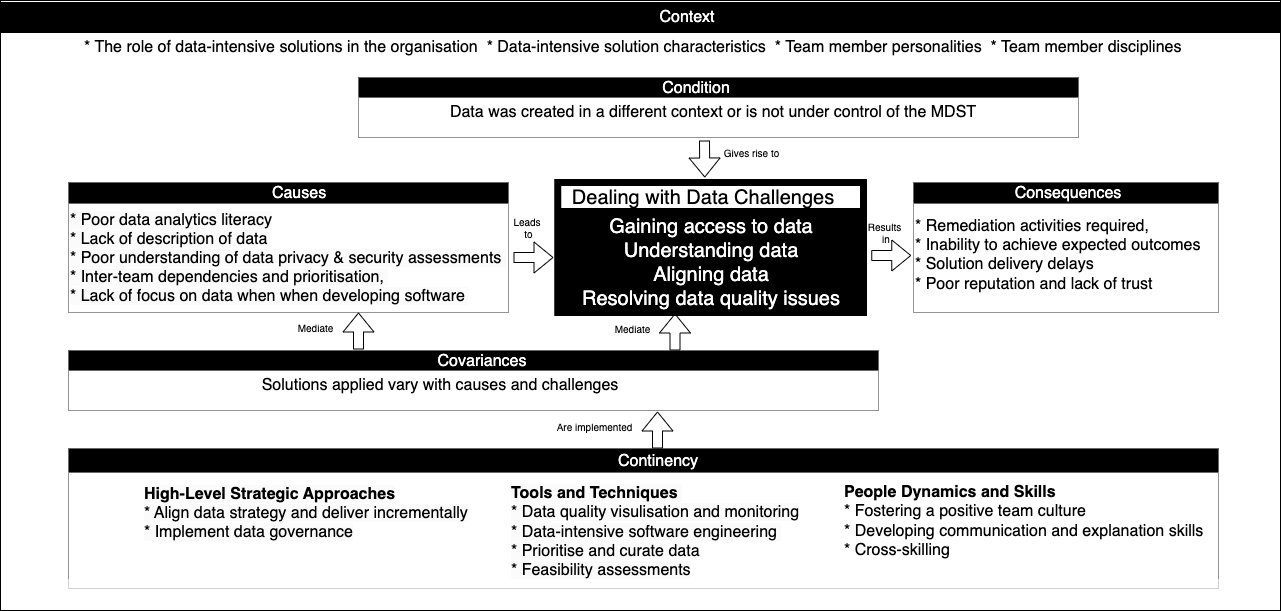}
\textcolor{black}{\caption{Theory of MDSTs dealing with data challenges when delivering DI software solutions structured and visualised using the 6Cs model \cite{Glaser1978TS}.}}
\label{fig:theory}
\end{figure*}


\section{Findings}
\label{sec:findings}
\textcolor{black}{The following sub-sections present \textcolor{black}{the results of our analysis structured into the} 6Cs model as shown in Figure \ref{fig:theory}. Our theory about multi-disciplinary teams dealing with data challenges explains: (a) the challenges faced by MDSTs including gaining access to data, aligning data; understanding data, and resolving data quality issues, 
(b) the context in and condition under which these challenges occur,
(c) the causes that lead to the challenges,
(d) the related consequences of these challenges, 
(e) the contingencies or strategies applied to address the challenges including high-level strategic approaches, implementing new tools and techniques, and building stronger teams,
(f) the relationships between strategies and causes and/or challenges represented as co-variances. Each sub-section describes one the 6Cs and includes pertinent quotes from the interviews that represent raw data used to define concepts and that highlight constituent properties.} 


\subsection{Context}
\label{context}
STGT studies and resultant theories are specific to their context. Our study's context was industry practitioners working in a mix of commercial and government organisations delivering different types of DI solutions.  Solutions included a vehicle telemetry system, a natural language processing analytics tool, business and operational planning and forecasting tools, business intelligence and analytics solutions, real estate applications with predictive capabilities,  and financial and health analytics solutions.  Interviews largely discussed experiences about systems that had been productionised -- only 2 systems did not yet have a production version and 1 system had failed. This section outlines four key contextual concepts distilled from our participant responses.

\subsubsection{The Role of Data-Intensive Solutions in the Organisation}
We found that experiences of challenges and solutions differ depending on the role that MDSTs and DI solutions play within an organisation. Specifically, this was due to whether participants were part of a team in an organisation that was in the business of developing DI related products/services, or whether they were part of a team that delivered DI solutions to support business activities. Where DI software development was not a key organisational objective driving operations, multi-disciplinary teams delivering DI solutions are considered a back office function. As a supporting function they have to adjust their communications and activities to align with their organisational 'frontoffice' context and goals. See Table \ref{tab:participants}, Software Org, which identifies if the organisation's predominant product is \textcolor{black}{software or not.} \par
 \faComment \hspace{0cm} \textit{"So when someone works for a tech giant, their product is their software. And analytics go hand in hand - that's the product they deliver. In our organisation, the product is the accounts, and related services, and the heroes are the branch staff, the frontline people, they are the ones dealing with the customer. It changes the dialogue."} P20 (Lead Data Analyst)

\subsubsection{Data-Intensive Solution Characteristics}
Key DI solution characteristics were discussed by P05, P08, P11, P12, P14, P17, P18 and P19.
When looking through the lens of traditional software engineering, DI solution delivery has some unique characteristics compared to the delivery of traditional software. Typically software development projects are described as feature driven and product focused. DI solutions typically combine data from different data sources which has been collected over long periods of time, has different owners and provenance, and different levels of quality and quantity.  DI solutions use models to generate insights which are the the system's result or  ``output".\par
\faComment \hspace{0cm} \textit{"Data analysis shows the results from the data. So it's not really that you have something made when you're doing something like a software solution."} -- P05 (Data Scientist)\par
\faComment \hspace{0cm} \textit{"You know, in the traditional software engineering space, we are very focused on a product"} P19 (Data Developer). \vspace{1mm}\newline
  DI solutions are exploratory and iterative.  Not achieving a result or changing direction is common.\par 
\faComment \hspace{0cm} \textit{"So once we were able to validate that it wasn't feasible, we had to kind of shelve the whole thing"} -- P12 (Technical Lead)\vspace{1mm}\newline
DI solutions also require continuous improvement to remain relevant, and cannot simply be administered like many traditional software systems.\par
\faComment \hspace{0cm} \textit{"The nature of data science work is that it always needs improvement – accuracy, parameters, and there are features where we made assumption and then generated the output.  Ongoing operations require more work and automation."} -- P11 (Data Scientist)

\subsubsection{Team Member Personalities}
Within MDSTs, the concept of predominant team member personalities in different roles emerged and were mentioned by P02, P04, P06,P08, P11 and P13 
Technical team members need to focus and work uninterrupted. Software developers and data scientists were both described as more introverted. Introversion and the inclination of technical team members to keep knowledge to themselves was identified as particular personality traits that impacts team communication.\par
\faComment \hspace{0cm} \textit{"The majority of our data professionals are extreme introverts...they do their job really well, but the handover and sharing doesn't happen."} -- P20 (Lead Data Analyst)\par 
\faComment \hspace{0cm} \textit{"I have worked with a developer who doesn't want to talk at all. He just wants to code. And he'll never tell you when he has finished."} -- P06 (Product Manager) \vspace{1mm}\newline
On the other hand, team members performing roles of business analysts and domain experts are described as extroverted, sometimes dominating.\par 
 \faComment \hspace{0cm} \textit{"Business Analysts tend to be extroverted, they can talk really well and influence outcomes."} -- P10 (Lead Data Analyst) 
 \vspace{1mm}\newline
Perfectionism and being passionate about their speciality could also impact teamwork because of their need for perfection which limits the ability to compromise.\par 
 \faComment \hspace{0cm} \textit{"Software engineering is my craft.  So I don't compromise much. You have to do things the right way."} -- P13 (Development Manager) \par

\subsubsection{Team Member Disciplines}
Team members' native discipline and its impact on communication and approaches was discussed by P01, P02, P04, P06, P18, P20 and P19.  Team members in MDSTs bring the knowledge, skills and perspectives of their respective discipline such data/software engineering, data scientists, finance, supply chain or other domain experts, each of which has its own language and approaches to delivering solutions. Participants highlighted the effort required by all team members when communicating and to clarify understanding with team members that have different backgrounds. Even team members that come from related disciplines, such as data science and information technology, may have a completely different understanding of terminology and domain knowledge.\par
\faComment\hspace{0mm} \textit{"Everyone's done different courses at uni. They've done different jobs leading into this, and so even a simple word like, ''data mode''l or''model'' can mean actually physically very different things to someone who's a data warehouse, or a data engineer, or a data scientist, or the domain expert or even the end-user."} -- P04 (Data Scientist)\vspace{1mm}\newline
Domain experts also find that team members from technical disciplines have more literal style of communication.\par
\faComment\hspace{0mm} \textit{"Particularly coming from a non technical background, it's almost like a different language in the way things are talked about. Some styles of communication, particularly on the technical side, I find they're very literal."} -- P18 (Consultant) 

\subsection{Condition}
\label{condition}
DI solutions rely on data, not just algorithms or programs to deliver an outcome or result. \textcolor{black}{Based on the interviews, data may be} sourced from disparate sources not under the direct control of the MDST.  Example sources include external data providers, legacy systems or historical databases. This characteristic has \textcolor{black}{been identified as a condition under which it is more likely that MDSTs have to "deal with data challenges", our key category.}\par
\textcolor{black}{
\faComment\hspace{0cm} \textit{''You are at the end station of where all the data comes.  And if the data is wrong at the source, you can't go back to the source...at our client, for example, they had to deal with data from businesses they had bought and sold'.'} -- P23 (Senior Manager)}\par
\textcolor{black}{
\faComment\hspace{0cm} \textit{''Data sources that feed us are from external data providers and they are not perfect and require a lot of processing but still bad data slips through."} -- P12 (Technical Lead)}

\textcolor{black}{\subsection{Data Challenges}}
The key category identified through the STGT analysis was `\textit{Dealing with Data Challenges}'. These include socio-technical data challenges including: \textit{Gaining access to data}, \textit{Understanding data}, \textit{Aligning data}, and \textit{Resolving data quality issues}. We found these challenges to occur in the context and under the condition described above in sections \ref{context} and \ref{condition} respectively.

\subsubsection{Gaining Access to Data} 
When MDSTs try to access data from sources that have not been used for similar purposes before, navigating procedures to gain access can be time consuming and challenging. These challenges can require navigation across organisational units - for example the IT Security team, the legal/privacy team, or any other team that is a designated 'data owner' and the engagement processes for those teams  are often poorly understood. This concept was articulated through discussions with P01, P04, P05, P08, P16, P22 and P24. Privacy and security related approvals may be required before data can be used for analysis. \par
\faComment\hspace{0cm} \textit{''Data security is a big issue for our organisation. When you want to get access to data through analytics apps we have to be sure that that whatever we're exposing is secure, and that the owners of the data are identified and they're okay with access to data.''} -- P22 (Analytics Architect)
\vspace{1mm}\newline
  The requirement to specify a particular purpose of use for the data can be quite a challenging requirement for exploratory analytics.\par
\faComment\hspace{0cm} \textit{''It's hard to tell [the data owner] what you want to do. Because you'll make a request for data, and they will say, 'Well, why?' And your answer is, 'I don't know yet'. You don't know what you can do with it.''} -- P24 (Senior Manager) \vspace{1mm}\newline
At a technical level, data and software engineers typically focus on navigating technical approvals and access protocols for the different data sources.\par
 \faComment\hspace{0cm} \textit{''There are challenges in  connecting into [client] systems. You know, like firewall restrictions, and  security implications of opening up your network to outside.''} -- P16 (Software Engineer)

\subsubsection{Understanding Data}
Challenges regarding understanding data were predominantly articulated by data and domain experts, specifically participants P02, P07, P08, P11, P14, P15, P17, P20, P21 and P22. When data used by the DI solutions is created in a different context, data analysts need to perform analysis to get an understanding of the data, including getting an understanding of the context of generation, what the data means and how it can address the problem or question at hand. Information to perform this analysis may not be readily available to MDSTs and time needs to be spent to find people and resources to gain understanding. They may also gain knowledge over time through data exploration.\par
 \faComment \hspace{0cm} \textit{``So the database is massive. One person doesn't know everything and we don't have contacts for all the data.  There were some projects that I said no to at the start - that I can't do this, that this isn't possible because we don't have enough data for that. Looking back at them, I now understand the data and I can provide a better solution.''} -- P08 (Data Analyst) \vspace{1mm}\newline
Data analysts may need to communicate with software and data engineers, end users and domain experts, and reconcile multiple perspectives to understand the data.\par
  \faComment \hspace{0cm} \textit{''I would go back to the engineer and then the database owner, and then I had a session with the customer who would say 'no I know the data, that should look like this, so you are making mistake'.  And I was back and forth getting people's feedback}. -- P11 (Data Scientist)\vspace{1mm}\newline 

Understanding complex domain data is subject to understanding the domain specific languages that can take years to learn. In highly specialised complex domains, such as medical specialities, domain expertise is costly and scarce. In less complex domain such as finance, challenges still exist -- for example there are different ways of calculating common metrics such as 'net sales' and it requires access to contextual domain knowledge to clarify. 
  
 \subsubsection{Aligning Data}
Interviews with data and domain experts P09, P10, P17, P18, P22 and P23 identified data alignment from different, incompatible sources as challenging. Examples of data alignment includes alignment of free-text to codes (to establish truth values for supervised learning), aligning data from different sources that have applied different categorisation scheme, or mapping data that has different timeframes.  Alignment here does not just refer to the technical aspect of aligning data - it also includes  the creation of mapping or alignment rules, processes to achieve and manage the rules and required stakeholder consultation. \par  
\faComment \hspace{0cm} \textit {``But the source data to create one curator table in Hadoop exists in three separate data platforms. Some are daily data, some monthly data, so they don't even have the same frequency.''} -- P10 (Lead Data Analyst) \vspace{1mm} \newline
Technical solutions such as mapping processes can be employed, but sometimes efforts to align data are abandoned because legacy units of measures cannot be converted.\par
\faComment \hspace{0cm} \textit{"Especially big companies who have many sites or plants where each plant has its own measures, like in one it's a metric ton and another its US ton. And then you can't sum those metrics together, because you have to do some conversion. In others they have compare bags and cartons and you can't compare those."} -- P23 (Senior Manager)\par
Where data mapping is used, solutions often offer limited point in time solutions and the mappings generally represent a particular stakeholder group's perspective.  
Additional complexities can be experienced when data has been entered free-text and multiple domain experts need to be consulted. Reconciling different perspectives can consume significant manual effort.\par 
 \faComment \hspace{0cm} \textit {``They can't even align names that we use to find the regions of interest on the scans. It's hard enough when you do it for people in your own department but when you come to another department -- it's mind numbingly hard.''} -- P17 (Oncologist)\par

\subsubsection{Resolving Data Quality Issues}
Serious data quality challenges, such as missing or invalid data were articulated by many participants: P01, P04, P05, P06, P08, P10, P11, P12, P17, P23 and P24.\par
\faComment \hspace{0cm} \textit {``There are definitely data gaps. And you know, it's not helped by the fact that we have so many legacy systems and platforms.''} -- P10 (Lead Data Analyst) \par
Whilst there are accepted approaches to dealing with data quality challenges including statistical options, their applicability needs to be worked through and may not be acceptable or available in the context of the project. \par
\faComment \hspace{0cm} \textit {``At one point in time, we had about 25 $\%$ of the data complete, and $75\%$ not complete, and that is a good data set...tell me what statistical technique do you have that will impute that data accurately?''} -- P17 (Oncologist)\vspace{1mm} \newline
Fixing such data quality issues can be challenging and cause tension for teams if they need to rely on other teams or team members to fix the issues.\par
\faComment \hspace{0cm} \textit{''There were lots of missing values i.e. features are missing but they were used to link data concepts. The data engineer on the team said “this is what I have” and then did the transformation and left some of the data out. When I did the model we found that if we want to infer something then we need those features.  I had to convince them. The data was very raw.  But you need all the pieces to make it work.''} -- P11 (Data Scientist)\par

\subsection{Causes}
The following subsections describe the identified causes that lead to \textcolor{black}{MDSTs having to deal} with data challenges. The identified causes were \textit{Poor data analytics literacy}, \textit{Lack of description of data},   \textit{Poor understanding of data privacy and security assessments}, \textit{Inter-team dependencies and prioritisation} and \textit{Lack of focus on data when developing software}. 
\subsubsection{Poor Data Analytics Literacy}
Poor data analytics literacy was identified by the participants across many disciplines: P03, P04, P05, P07, P08, P11, P12, P15, P17, P18, P21 and P23.
Low level of understanding of data analytics concepts and DI solution characteristics within the MDST or wider organisation was identified as limiting \textcolor{black}{the understanding of the data, 
and the pace at which issues can be resolved. More knowledgeable members of the team end up having to explain concepts and educate and take on the burden of dealing with the data challenges as articulated in Section 4.3} \par 
\faComment \hspace{0cm} \textit{"
In teams like this, where the team doesn't understand the data, or the team doesn't understand the data concepts...it is a bit of a challenge to get them to understand what we can actually do with the data itself."} -- P08 (Data Analyst) \vspace{1mm}\newline
Junior practitioner interview participants, skilled in data analytics described experiences where more senior team members, including those in leadership decision making position had limited understanding of data analytics.\par
\faComment \hspace{0cm} \textit{"When I joined the team, I had to educate the Project Manager, I had to educate the customer, I had to educate my team members and our manager. So none of them had any clues about the concept of a data science project. Data Engineers were purely focused on the raw data so they had no clue about what I wanted to do."} -- P11 (Data Scientist)\vspace{1mm} \newline
\textcolor{black}{The lack of data analytics literacy is also experienced as very limited use of data project methods in use when MDSTs deliver DI projects. Whilst data expert participants and software engineers mentioned the use of agile methods in their project overviews, only one participant articulated the use of a data project process method.}\par
\faComment \hspace{0cm} \textit{"I have been trained through Microsoft Azure which taught us a process and I developed them further with my manager."} -- P05 (Lead Data Scientist)
\vspace{1mm} \newline
\textcolor{black}{Lack of data analytics literacy can manifest as lack of awareness by those creating the data about the potential use and value of that data to DI solutions. The lack of understanding or considerations for future data-intensive solutions that consume data can lead to MDSTs having to deal with alignment issues and resolution of data quality issues as discussed in sections 4.3.3 and 4.3.4.}\par
\textcolor{black}{\faComment \hspace{0cm} \textit{"Training for doctors is zero...most doctors say, 'Why am I entering this? What will happen with this information?...they don't understand that our routine data can be used for data mining"} -- P17 (Oncologist).}
\subsubsection{Lack of Description of Data}
Participants P08 and P15 articulated the meaning and context of data generation is not documented and this causes downstream challenge \textcolor{black}{of not understanding the data}. While analysts have access to available documentation, table structures, names and column names, they typically do not have access to the data entry screens or context of the business process that generates the data.\par
 \faComment \hspace{0cm} \textit{''And the challenge, for the [Data Analyst] was the implementation of the field by the engineer, actually stored in the database, as actually, you know, it's named differently to how it's serviced to the UI and to the, you know, documentation of these piece of functionality...\textcolor{black}{And the data scientist, without access to code, or the implementation couldn't understand by looking at the database field.''}} -- P15 (Lead Engineer)

\subsubsection{Poor Understanding of Data Privacy and Security Assessments}
Participants P03, P04 and P05 described lengthy security processes and privacy assessment protocols that delay projects. The privacy assessment protocols are enshrined in legislation and form part of organisations compliance regimes. \textcolor{black}{In turn, this leads to teams having to deal with data access challenges such as described in section 4.3.1 and some work not progressing at all due to legislative constraints}.\par
 \faComment \hspace{0cm} \textit{''The user story says that the patient's history should be displayed. So that went into the user stories. The legislation, the privacy, the Human Rights charter, all of these things prevent us from having access to that information. So it was never going to be achievable. Yet, that was one of the user stories that went into the program of work, and so they built a system and built it into the dashboard, but could never get the data feed or the information.''} -- P03 (Director of Data) \vspace{1mm} \newline
Even if the approval processes are understood, the iterative nature of DI Solutions may identify need for data progressively and require multiple rounds of approvals and assessments. Poor understanding of these requirements, and lack of allowance for those processes can cause \textcolor{black}{repeated} delay to data access.\par
  \faComment \hspace{0cm} \textit{''And later, we had to get clearance for more data fields. So we had to go through the whole process of approval for data again.''} -- P05 (Data Scientist).\par
 
 \subsubsection{Inter-Team Dependencies and Prioritisation}
Inter-team dependencies and prioritisation of work were described by several participants, including P01, P04, P05, P08, P21 and P22. It was common for MDSTs to depend on other teams to get data quality issues fixed (see section 4.3.4), system functionality fixed or questions answered.\par
  \faComment \hspace{0cm} \textit{''When collaborating with the data engineering team or data warehousing team, we do have to put out a request for them to fulfill to get extracts from them, or getting some piece of data from them, which we can then modify. Sometimes the waiting period of this as an issue.''} -- P08 (Data Analyst)\vspace{1mm}\newline
The other teams have their own priorities or may not have the capability to provide the solution.\par
\faComment \hspace{0cm} \textit{"We might say that we need this data, or that there's a bug or a particular set of data are not coming through properly. Whilst that doesn't actually matter to any of their downstream processes, it matters to our processes. Can you fix it please? And it's like "we have other priorities".} -- P01 (Data Architect)  

 \subsubsection{Lack of Focus on Data when Developing Software}
Limitations of existing software engineering practices and skills were discussed by participants P01, P06, P13, P19, P21.   
  One limitation is that data-intensive solutions may need to analyse data generated in a different context, by a solution or system that was developed without consideration of the needs of future downstream data-intensive solutions.   For solution that rely on human data entry, data entry standards which are key to achieving high data quality, are not sufficiently addressed during source system development. \textcolor{black}{In turn this leads to teams having to deal with data challenges, specifically those discussed in sections 4.3.2. and 4.3.3.}\par 
\faComment \hspace{0cm} \textit{"They don't really prioritize reporting and analytics when they build the source systems. And it manifests into things like for example, if I do want a column to be highly structured, and the quality of the data in that column is very high, you usually want it to be a drop down. So the user can only select from certain values. When people run out of time, they will just go with a free text column. And this sort of thing causes a lot of havoc downstream."} -- P21 (BI Manager)\vspace{1mm} \newline
Software engineering participants expressed interest in data frameworks and architecture to ensure a sound footing and foundation of the solution. However, they rely on domain experts to articulate features that end users want and are not necessarily concerned with the actual data being managed by the system.\par
\faComment \hspace{0cm} \textit{''So we're making the app for his team. And so the domain expert will tell us essentially what the financial end user wants. What charts, what tables and data,you know, all these finance items. For me, they mean absolutely nothing.''} -- P13 (Development Manager) \par

 \subsection{Consequences}
Having to deal with data challenges led to the following adverse consequences: \textit{Remediation activities required}, \textit{Inability to achieve expected outcomes}and \textit{Solution delivery delays}.  Flowing on from these are \textit{Poor reputation and lack of trust} as additional consequences.   

\subsubsection{Remediation Activities Required}
A direct impact of poor data quality in DI applications is that there may be legal compliance related reasons that demand remediation. Remediation can be far reaching and require redesign and re-implementation of the solution to address how data is processed or to address privacy concerns.  This impact was raised by participants P17, P21 and P10. Remediation may involve data mapping or data cleansing activities.\par
\faComment \hspace{0cm} \textit{''We often have to do manual cleaning as well. Either we go into the source and fix the gap in the spaces or we just run that into a Excel sheet and then do a mapping.''} -- P21 (BI Manager)

\subsubsection{Inability to Achieve Expected Outcomes}
Participants P03, P21 and P22 discussed that issues with data quality makes it difficult to analyse data and may result in solutions that do not meet the required predictive accuracy.  Lack of knowledge about what data is available can prevent solutions from being conceived in the first place. Lack of knowledge about data quality and simply performing analysis on data that is available, can lead to low value outcomes.\par
  \faComment \hspace{0cm} \textit{``They advocated machine learning would be possible with the product that they were building. But the poor data quality meant that the end product is 20\% of what they initially set out to achieve, and therefore makes it largely not fit for purpose.''} -- P03 (Director of Data) 

\subsubsection{Solution Delivery Delays}
Solution delivery delays due to data challenges were mentioned by P05, P11, P21, P22 and P23. Some of the causal factors such as poor data analytics literacy exacerbated the challenges and slowed down delivery. Teams working together for the first time can spend significant amounts of time working out how to work together to deliver a DI solution. Data approval processes and inter-team dependencies add to the delays.\par
  \faComment \hspace{0cm} \textit{``One of the challenges is timelines, mostly due to working with the [client organisation], they need a lot of time for decision making approval of which data fields can be used.''} -- P05 (Data Scientist)\par
  \faComment \hspace{0cm} \textit{``And, you know checking that the owners of the data are ok with access to data. All of this adds up to the duration and the cost of delivering. What may seem simple, ends up taking a long time to do and can be quite expensive.''} -- P22 (Data Architect)
  
\subsubsection{Poor Reputation and Lack of Trust}
Participants P03, P17, P23 and P24 discussed the downstream consequence of poor reputation and lack of trust resulting from dealing with data challenges.  This could be due to performing some of the remediation activities that are  viewed suspiciously and lead to  outputs from data analytics solutions not being trusted.\par
\faComment \hspace{0cm} \textit{''My suspicion is that it will work on your data in your center and nowhere else...I don't trust anyone who's sending me a machine learning algorithm.''} -- P17 (Oncologist)\par
  \faComment \hspace{0cm} \textit{''And then you typically do some cleansing, or some like alignment. But whilst that is workable people have issues with reconciliation, because you sort of  start mapping and changing data, in essence, and then people start spending time on reconciliation to make sure it matches, and you have a lower trust factor.''} -- P23 (Senior Manager)
 \vspace{1mm} \newline
Ultimately this reflects poorly the reputation of the solution and the MDST that created the solution.\par
  \faComment \hspace{0cm} \textit{''That can either result in a delay in the delivery time, or a bad reputation. Often they say it is the solutions team doing the poor job.''} -- P21 (BI Manager)

\subsection{Contingencies}
Participants provided insights into the strategies and techniques that they are actively using or trialling to address data challenges in their teams. 
We grouped the approaches into \emph{high-level strategic approaches},  \emph{tools and techniques} and \emph{people dynamics and skills}.  

\subsubsection{High-Level Strategic Approaches}
Executive level participants articulated longer term (multi-year), overarching strategic objectives and approaches to dealing with data challenges including:\par
\vspace{3mm}
\noindent \textit{4.6.1.1 Align Data Strategy and Deliver Incrementally.} At the executive level, the focus is on designing data strategies and the need to deliver DI solutions in parallel with  associated infrastructure. Data frameworks and data quality on their own are not recognised to have sufficient intrinsic value. There is a need to design projects and initiatives so they align with the data strategy and each initiative needs to deliver tangible benefits within funding cycles. This approach was discussed by participants P02, P03, P14 and P22.\par
\faComment \hspace{0cm} \textit{``Data quality and data integration are fairly dry subjects to [Exec Board], because they can't materially see the end point. If you're trying to establish a benefit, particularly a customer benefit, [Data Framework] sort of work doesn't stack up. You are not getting the strong customer outcomes... So rather than bite it all off in one big piece, I've broken it up into smaller pieces, to make the task more achievable.''} -- P03 (Director of Data) \par
\vspace{3mm}
\noindent\textit{4.6.1.2 Implement Data Governance.}
Implementation of data governance came up as an approach for dealing with data challenges in discussions with P02, P03, P14, P19, P21 and P24.
Executive level participants were in the process of implementing multi-year data governance initiatives.  When asked for more details, it became clear that this term does not have a consistent definition or imply the same priorities.  For one executive, the priority was to address data literacy, drive data quality, and ensure people understand their role in relation to data.\par
\faComment \hspace{0cm} \textit{``Establishing data governance, and getting a terms of reference in terms of their domain, and then trying to build up literacy to get to a point where people can understand what a data product is, what their roles is relating core organsiational master data and data management in general''} -- P03 (Director of Data)\vspace{1mm}

Master data here refers to the core data that an organisation needs to support transactional applications including customer data, suppliers, products, account codes etc. For another executive, the priority was to develop a common domain language across different organisations and clarify what should be common and what can remain flexible/separate.\par  
\faComment \hspace{0cm} \textit{``So hopefully, we can use a common language format using data governance layer or data sovereignty layer to solve this problem without touching the technical details.''} -- P14 (Chief Information Officer) \par
For P02, General Manager Data Analytics, the goal was to ensure their teams have the right tools and support to deliver the data strategy.  On the other end of the spectrum, BI Manager P21 saw the implementation of formal data governance as something for large organisations only, and his organisation of 300 employees was too small. 
P19 (Software Engineer) mentioned that `hardening' data quality had now been included as a team performance measure. Whether data governance initiatives go far enough to achieve their objectives is not clear -- even organisations that have introduced data governance roles, struggle to articulate data quality responsibilities.\par
\faComment\hspace{0cm} \textit{``Some of my clients have clear data owners for specific fields and datasets, and they might own it -- but how much responsibility do they have over filling in the gaps? Depends on what it's used for.``} -- P24 (Senior Manager)\par

\subsubsection{Tools and Techniques}
Participants identified a number of tools and techniques that their teams were using to address data challenges:\par

\vspace{3mm}
\noindent \textit{4.6.2.1 Data Quality Visualisation and Monitoring}
Some software engineer participants (P15, P19) discussed their setup and plans for developing tools, whilst some data analysts and domain experts discussed their use and limitations of tools (P21, P18, P20, P21) to analyse data quality.\par 
\faComment \hspace{0cm} \textit{``We collect the data and we pipe it through to a data lake and visualise those data on a dashboard. So the tool serves up data quality. That tool was used to build an application for purpose, so that it's a tool to understand what is missing, what data needs to improve itself.''} -- P15 (Lead Engineer)\vspace{1mm} \newline
While the tools assist in visualising and analysing data, they rely on custom development or configuration to analyse data quality, which needs to be targeted to the specific domain context to ensure that any alerts are relevant.\par
\faComment \hspace{0cm} \textit{``To make these monitoring tools really useful requires extensive domain knowledge, because you've got to know what sort of ranges you're looking for, because otherwise you will be bombarded with a lot of alerts and notifications which are not relevant.``} -- P21 (BI Manager)\vspace{1mm} \newline
Data quality monitoring tools are being developed and used to make data quality continuously 'observable'.\par 
\faComment \hspace{0cm} \textit{``We want see what's running through our pipelines a lot more easily. So accessing, producing dashboards and sort of stuff to visually identify where anomalies might occur. And if data is going in at one point, does it flow through.''} -- P19 (Data Developer)\par
Some constraints on implementing tools are time and cost -- the tools take time to learn and implement and may have ongoing usage costs.\par

\vspace{3mm}
\noindent\textit{4.6.2.2 Data-Intensive Software Engineering.}
Participants P01, P06, P14, P19 and P21 articulated that more data-intensive approaches to software engineering can help prevent or reduce data challenges \textcolor{black}{in} future DI solutions. This includes designing DI solutions to be more self documenting and implementing frameworks such as Data Build Tool (DBT) to future proof DI solutions.\par
\faComment \hspace{0cm} \textit{``A YAML file that has other metadata in it as well, that allows for documentation and allows it to be viewed as code for running, but also viewed as documentation for people actually doing the data analytics.''} -- P01 (Data Architect)\vspace{1mm} \newline
Involving analytics experts within requirements engineering and initial design stage can also influence data quality design for future analytics reporting and lead to better downstream understanding of the domain processes that generate data for analytics.\par
\faComment \hspace{0cm} \textit {``I will have a better opportunity to interact and decide things like, what sort of transaction types are we going to have. What will our Chart of Accounts look like.  How do you reflect the business process..because the knowledge of how the data is generated in the first place is extremely important for my work.``} -- P21 (BI Manager) \vspace{1mm} \newline
Another approach when developing new systems is identification of relevant, valuable legacy data and enabling new systems to absorb this.  Over time, this will enable decommissioning of legacy systems and contribute towards providing a centralised trusted sources of data.\par
\faComment \hspace{0cm} \textit {``And the reality is, you have to keep the old platforms running, and then at same time, you build new systems to absorb the data. Instead of consolidating the systems, we consolidate the data''} -- P14 (Chief Information Officer)

\vspace{3mm}
\noindent\textit{4.6.2.3 Prioritise and Curate Data.}
Participants P10, P20 and P23 discussed better curation of data as an approach to address many data challenges. Specific datasets are prioritised and then effort is spent to curate data before it is analysed.  Curation includes cleaning the data, and performing checks -- including fit for purpose testing by domain experts.  The challenge with this approach is that it is time consuming and expensive, hence the need to prioritise.  Typically the data is curated for a specific (prioritised) single purpose only.  Where multiple domain experts are required to curate or test data, consistency issues can emerge. For some projects, timing and context can play a factor in whether curation/data cleansing is as suitable solution. If data quality issues are identified early and their impact is considered significant enough, then special data quality projects can be initiated.\par
\faComment \hspace{0cm} \textit{``Data, sanity checks, and correctness, and hygiene. The one thing that we've also started doing, very actively actually, is getting Squads that are going to be customers of these data sets to actually do ``fit for purpose testing'' on the data.''} -- P10 (Lead Data Analyst)\par

\vspace{3mm}
\noindent\textit{4.6.2.4 Feasibility Assessments. }
Feasibility assessment were mentioned by P22 and P23. Prior to starting a new DI solution, and assessment is performed to assess data quality and estimate effort to secure data approvals. 
The  assessments identify data quality deficiencies and data approval dependencies and may result in a decision for initiative to not go ahead.
\par
\faComment \hspace{0cm} \textit{''But all of this adds up to the the duration of the project, and the cost of delivery.  What may seem like a simple request ends up being estimated to take a long time because of a lot of approval and alignment processes you need to go through to to deliver that, and it can be prohibitively expensive.  Then people may say, "Oh, don't worry about it''} -- P22 (Analytics Architect)\par

\subsubsection{People Dynamics and Skills} 
Working in MDSTs requires strong inter-personal skills.  Developing a positive team culture, cross-skilling, and communication and explanation skills were identified approaches for addressing data challenges and to mitigate data challenges.\par   
\vspace{3mm}
\noindent \textit{4.6.3.1 Fostering a Positive Team Culture.}
Participants P02, P06, P08, P09, P14 and P19 identified MDST and wider organisation culture as an important human-centric aspect in MDSTs for addressing data challenges. Team members that experienced positive culture made specific reference to the impact that has on their ability to develop good relationships with team members and beyond their immediate team to resolve questions and improve data understanding.\par
\faComment \hspace{0cm} \textit{''But our organisation is good in the sense that you can reach out to anyone in the business and you like, someone's gonna be like, Oh, that's cool. Let's have a chat or something like that.''} -- P19 (Data Developer)\par
Participants in leadership roles highlighted the importance of and their efforts in building an open and safe culture where collaboration and questions are encouraged. Techniques for building and fostering positive team cultures include facilitated social discussions and team quizzes. Building a positive culture takes effort and requires leadership by example and recruitment activities that select for cultural fit.\par
\faComment \hspace{0cm} \textit{``If you can have the right people you can have the right product discussion there. The next things is you build the best team culture, let them run to do their part, I'm less worried about technical details, to be honest. If you can have the right people and give them enough support, they will solve the problem for you.''} -- P14 (Chief Information Officer) \par

\vspace{3mm}
\noindent \textit{4.6.3.2 Developing communication and explanation skills.}
The need for excellent communication and explanation skills was discussed by many participants: P1, P2, P3, P4, P5, P7, P8, P9, P10, P11, P15, P17, P18, P19, P20, P21 and P24. Working in MDSTs requires a high level of patience, strong inter-personal skills, a willingness to seek clarification and foster understanding. Overall, excellent communication and explanation skills about data analytics concepts can help bridge the gaps in data analytics literacy, facilitate data understanding and clarify what solutions can and cannot achieve. Effective communication and explanation skills require an understanding of the audience and their domain, an ability to translate between data and domain/technical concepts and the ability to break down and simplify concepts.\par  
\faComment \hspace{0cm} \textit{So what we try to do is to to translate our prediction towards a daily result for them. So how many people will arrive in the hospital? How many people  will be waiting longer than we predict? ... and is that something you can live with, or not? And then we'll discuss what would happen if we have say 10\% of People who might wait longer? How will you manage those people? And how does that feel for you? We try to really discuss the consequences of such a prediction.''} -- P05 (Data Scientist) \vspace{1mm} \newline
Explaining concepts through a combination of verbal and visual tools such as diagrams, domain models and visual data modeling tools can be useful.\par
\faComment \hspace{0cm} \textit{''I like to use diagrams such as domain models. Diagrams are easy for people to understand, this allows focus on looking at one thing and understand the language we're talking about.''} -- P15 (Lead Engineer)\vspace{1mm} \newline
Approaches that worked well for participants to develop communication skills include being coached and getting supported by a more senior team member in day to day meetings, and attending business language courses.\par
\faComment \hspace{0cm} \textit{''I have been working there for a year and a half. Now, I understand how to explain to non technical people in a simpler fashion. Based on what their expectations are out of data, basically just explaining the domain context of everything that I'm doing.''} -- P08 (Data Analyst)\par

\vspace{3mm}
\noindent\textit{4.6.3.3 Cross-Skilling.}
Cross-skilling between data, software and domain expert roles was discussed with participants P04, P05, P06, P08, P10, P11, P12, P17, P21, P22 and P23.  Cross-skilling of team members can improve understanding of each others roles and make it easier to manage resources and to fill skill-gaps. 
Cross-skilling technical team members in the domain is a common approach and can be achieved using formal techniques such as such as process mapping, data mapping and domain modeling.  However, predominantly, organisation specific domain knowledge is still gained through day to day working.  Cross-skilled developer, domain and data analytics experts are highly valued.\par 
\faComment \hspace{0cm} \textit{''To truly implement the best solution, and that means putting your fingers on the keyboard and actually developing them out, you need to be quite heavily involved in the requirements and the pain points to truly understand exactly what needs to happen. So in our world, we specifically hire people that are very well rounded, and then have experience in all areas and all facets''.}  -- P09 (CEO)\vspace{1mm}\newline
Rather than trying to recruit already cross-skilled team members, several organisations are providing opportunities to cross-skill at the technical level. Offering skills development such as SQL skills to domain experts as a means to self-serve when there are resourcing constraints worked for one team manager. For more advanced software engineering skills, this requires additional effort, such as implementation of `guardrails' and quality assurance processes to ensure quality outcomes are not impacted when a team member is building skills. Early indications are that trying to cross-skill data scientist team members with software engineering skills may be more challenging than expected.\par 
\faComment \hspace {0cm} \textit{``I do have like, data scientists trying to write code and you kind of expect them to kind of improve in a couple of months, but, everything is still a bit mediocre.''} -- P12 (Technical Lead )\vspace{1mm} \newline
Cross-skilling domain experts with data analytics skills is also desired, but not yet prevalent.\par
\faComment \hspace {0cm} \textit{''We should have probably the data scientist playing a bit more with accountants and there should be probably a bit more uplift of accountants to become a bit more statistical.''} -- P23 (Senior Manager ) \vspace{1mm} \newline
Bridging larger technical gaps such as skilling domain experts in software engineering is considered as too large a gap and cross-skilling domain experts by training them on technical tools without specific organisational context such as data and processes is seen as offering low value. There is also a need to create career paths and reflecting cross-skill achievements and goals in professional development plans.\par

\begin{figure*}[!h]
\includegraphics[width=1.0\textwidth]{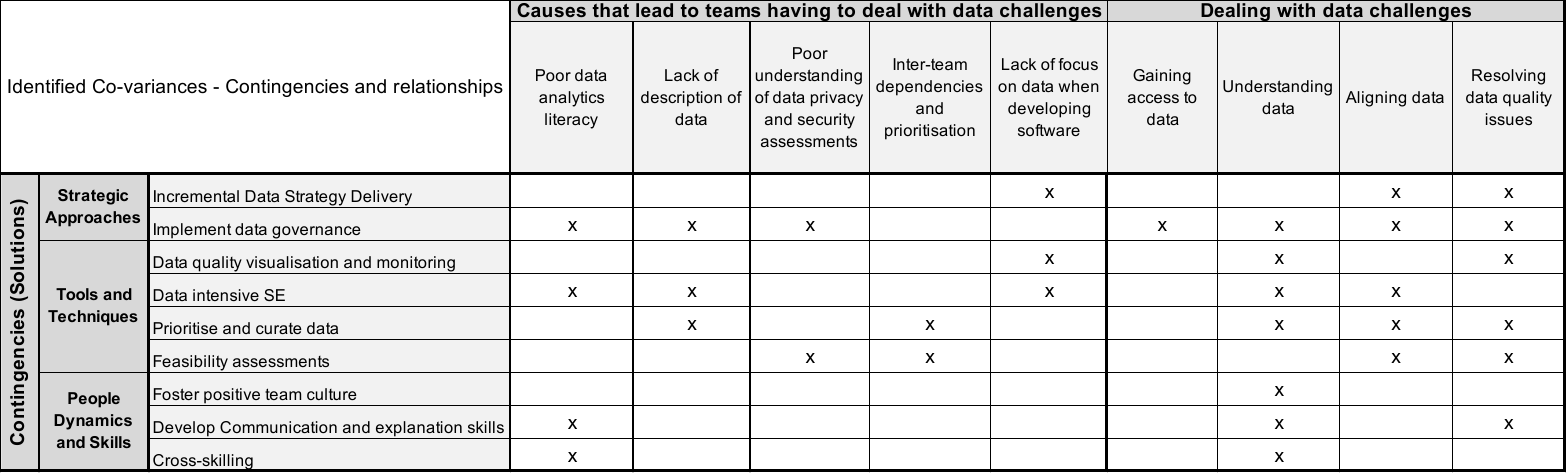}
\caption{\textcolor{black}{Summary of covariance relationships between Contingencies, Key Category concepts and Causes.}}
\label{fig:covariance}
\end{figure*}

\subsection{Covariances}
\textcolor{black}{Covariance occurs if a variable  changes when another variable changes\cite{Glaser1978TS}. 
Our analysis found data to support that the contingencies vary depending on the challenges and causes. Covariances are used to represent these relationships.  A summary of identified relationships is presented in Figure \ref{fig:covariance}. We present two examples to aid understanding. Examples in Section 4.6 are also relevant to this section.}\par  
\textcolor{black}{\emph{Focusing on data when developing software} is supported through developing and using \emph{data quality visualisation and monitoring} tools that observe data and transformation flows through pipelines. The development of textit{data quality visualisation and monitoring tools} supports MDSTs in \textit{resolving data quality issues}.  \textit{Implementation of Data strategy} is also focusing software engineering attention on data.
}\vspace{0.2cm} \par
 \faComment \hspace{0cm} \textcolor{black}{\textit{''We try to make sure our data flows through with a lot of observability in case issues come up, making sure that end to end data quality is very high...this has come through as a strategic objective for this quarter to monitor for critical data points and this is linked to our KPIs''.} --P19 (Data Developer)}\par 
  \textcolor{black}{Improving documentation in solutions through \emph{data-intensive software engineering approaches} aims to increase in \emph{focus on data when developing software} and seeks to address \emph{lack of description of data}} and improve \emph{understanding of data}\vspace{0.2cm}\par 
 \faComment \hspace{0cm} \textcolor{black}{\textit{''We hope to move towards DBT for doing transformations because... we want to be able to have documented processes, documented data products and understandable data products.''} --P01 (Data Architect)}

\section{Discussion}
\label{sec:discussion}
We now discuss our findings in relation to related works, as part of STGT's \textit{targeted literature review} procedure conducted in the advanced stages of theory development to ``\textit{compare the emergent original findings with existing work and situate them in the wider research landscape, filling research gaps}''. This is followed by recommendations for practitioners and researchers based on our  original findings and relevant ideas from related works.
\textcolor{black}{\subsection{Discussion of Findings in the context of relevant literature}}
\subsubsection{Data challenges and data quality} In 1996, Wang and Strong 
identified common patterns of data quality problems, including the need for time and effort to get authorised permission, interpretation and understandability of data, and lack of consistent representation as being barriers to access and use data. They identified that missing data and data coming from multiple distributed data sources make it difficult to integrate data and they also identified data volumes and limited processing power as challenges\cite{strong1997data}, \cite{wang1996beyond}. Our findings indicate that apart from data volume and processing capabilities, similar challenges still prevail more than 25 years later. Our study found that modern MDSTs are dealing with broad data challenges -- including \textit{understanding data} and \textit{gaining access to data}.  Domain experts are routinely and repeatedly required to work with MDSTs to understand and interpret data -- and sometimes the right domain experts are not available or expensive -- e.g. medical experts. Before gaining access to use data in DI applications, MDSTs have to seek approval from the appropriate data owner, adding more time and complexity to development processes and resulting in delays in solution delivery.  In addition, MDSTs also need to navigate challenges about data privacy.
Our study also found that when multiple data sources are combined, there continue to be challenges of \textit{aligning data} from multiple data sources, which is further complicated when there is also an \textit{understanding data} challenge. Ultimately these challenges result in solutions that do not achieve the required accuracy or outcome.\par
Research into data quality issues, costs associated with poor data quality, benefits to be achieved by good data quality and strategies to improve data quality continues to be topical \cite{wang1996beyond,CORTEREAL2020103141,gudivada2017data}. 
Additional challenges have been identified for \textcolor{black}{DI software systems} due to increasing use of big data, including the need to cater for data quality in high volume, high velocity and variety of data and the need to harmonise data \cite{7374131juddoo,MERINO2016123,gudivada2017data,10.1145/3394486.3406477Jain,ardagna2018context}. \textcolor{black}{Whilst} the concept of data quality and related data quality dimensions have been researched extensively \cite{scannapieco2005data}, there is still no agreed definition of data quality nor the dimensions that define it\cite{10.1145/1541880.1541883batini,8457745taleb,8642813cichy}. Gudivadada et. al\cite{gudivada2017data} proposed a data governance driven data quality life cycle framework to address issues exacerbated by big data, including new data quality dimensions to address machine learning requirements such as gender bias, confidentiality and privacy, access control, feature selection, feature extraction and outliers, but this has not been empirically evaluated. \textcolor{black}{Whilst our study participants did not specifically refer to any frameworks, we found that MDSTs are conducing \textit{feasibility assessments} prior to commencing initiatives and are seeking to assess some of these issues. There is a need to empirically evaluate frameworks.}\par
\textcolor{black}{The time taken to dealing with data challenges, specifically resolution of data quality related challenges continues} to top the lists of challenges cited by practitioners\cite{Hayes2018,Muller2019,Rattenbury2017}. Research is ongoing to find better solutions for dealing with missing values in domain specific contexts \cite{Ehrlinger2018}. However, some researchers argue that data issues are just part of data science work \cite{aho2020} and there are many tools and methods available for data wrangling and dealing with missing data values \cite{aho2020,Kandel2011,Rattenbury2017,Little2019}. %
%
\textcolor{black}{Our study found that \textit{resolving data quality issues}, including addressing missing values still} presented challenges in different domains, and one particular domain expert was very critical of the manner in which statistical techniques are sometimes applied and the conclusions that are drawn, leading to a lack of trust in the outcome of the solutions. \textcolor{black}{Additional research is required to bridge the seeming disconnect between practitioners who need to deal with data issues and the research perspective that the issues have been addressed. Empirical studies are required to determine if this lack of trust could be addressed through better data analytics literacy or explanations of how the missing values were addressed}.\par
Data quality has \textcolor{black}{also} become an important topic for machine learning practitioners\cite{10.1145/3394486.3406477Jain, gupta2021data} and they are looking to shift their focus from “goodness-of-fit” to focus on "goodness-of-data"\cite{Sambasivan2021}. More broadly than just data quality, industry leaders are calling for a shift towards a data ecosystem for industrial enterprises which will unite data producers' data sources with consumers through data application platforms\cite{10.1145/3448247groger}. \textcolor{black}{This last approach may offer some promise to address one of the causes identified in our study i.e. the lack of focus on data by those developing source systems. Our findings relating to \textit{implementing data quality visualisation and monitoring} are also consistent with this direction towards eco-systems.}
\subsubsection{The role of data governance in addressing data challenges}  
Data governance provides a cross-functional framework for managing data as a strategic enterprise asset. It specifies decision rights and accountabilities for data decisions, provides formalised data policies, standards, and procedures, and monitors compliance \cite{abraham2019data}. The ability for big data analytics capabilities to drive innovation, and the \textcolor{black}{positive} effect that data governance plays has been empirically confirmed
~\cite{MIKALEF2020103361}.
The positive relationship between data governance and data quality has also been explored empirically
\cite{al2022investigationWahshi}.\par  
However, a recent “practical investigation” of data governance structures in industrial enterprises 
 highlighted the rudimentary implementation of organisational structures to support data governance\cite{10.1145/3448247groger}. Often the data owners are the same as the source system owner, leading to multiple approvals required when multiple data sources are combined in a solution. Data stewardship is generally not centrally organised, leading to different policies, quality criteria and standards across data owners. 
Also, there are also still outstanding research questions about how privacy requirements can be addressed in a big data environment, appropriate data quality measures for big data, definition and measurement of intrinsic data value, and integration of data silos~\cite{abraham2019data}.\par
Our findings are partially consistent with the literature. Participants in \textcolor{black}{senior roles} considered data governance as an important long term strategic solution to address data challenges. Our study identified that participants each described their implementation and goals of data governance differently most likely due to the need for executives to align their vision to their own organisational strategies and structuring initiatives to deliver regular, tangible results. Their implementations were in progress and the participants did not discuss the results of the implementations. \par
Our study identified two instances where feasibility studies for new projects were conducted that involved consultation with data owners. In one instance, \textcolor{black}{data related feasibility considerations were included as part of the initial project feasibility assessment} and took into consideration the estimated time and effort required to obtain approvals by data owners.  This assessment led to the project being cancelled due to the projected timeframes. In the other instance, the feasibility assessment led to commencement and completion of a data quality remediation project before commencing the data-intensive solution. However, our study also found data governance does not in itself address the issue of resolving data quality issues. Those responsible may not have the resources to address the issues or the request for quality may be outside the anticipated scope. So, whilst data governance \textcolor{black}{holds out promise for the future, the different approaches to governance implementation and our participants' experiences suggest that it is currently offering mixed benefits at best}.\par 
Data governance \textcolor{black} {also has a role to play in uplifting general} data literacy and \textcolor{black}{in turn improve} data quality\cite{Koltay2016}. 
Data and analytics literacy is \textcolor{black} {considered} critical to realizing the value of commercial information assets \cite{Miller2014}.
Within our study, only some participants that discussed data governance implementation explicitly identified that data literacy uplift was a goal in implementing data governance. Our study \textcolor{black} found that poor data analytics literacy \textcolor{black}{was} a cause of data challenges. For example, if domain experts do not understand how their data is used in data-intensive solutions, they will struggle to make it fit for purpose. Our participants also identified that poor data analytics literacy made communication between data experts, domain experts and software engineers more difficult. \textcolor{black}{Given the relative low level of data analytics literacy reported by participants, data governance programs still have a long way to go before they deliver on their promises.}
\subsubsection{\textcolor{black}{Methods for data-intensive solutions}} 
\textcolor{black}{During our interviews, participants did not mention data project processes such as those based on the Cross-Industry Standard Process for Data Mining (CRISP-DM)\cite{crispchapman2000} which is still considered the de-facto industry standard\cite{Martinez-Plumed2021}. According to a Kaggle poll conducted in 2014, 43\% of participating respondents claimed CRISP-DM as their main methodology, making it the leading methodology, despite it being more than 20 years old\cite{Piatestki2014}. In that same poll, in second place, 27\% of respondents claimed to use their own.}\par
\textcolor{black}{However, in our study, merely one participant, who worked as a data science consultant, referred to an in-house developed data project process based on Microsoft processes. Our participants generally referred to Agile as the predominant way of working.  Domain experts did not provide reference to processes. None of our participants called out a formal data project method as either being used or being considered. This lack of use of data project methods aligns with findings by Saltz et.al.\cite{saltz2018exploring} who conducted a survey of data science teams and the project management methodologies in use as of 2018, where 82\% of respondents did not follow an explicit process.}\par
\textcolor{black}{From a software engineering perspective, the growing importance of considering data science and its processes is evidenced by the call from the IEEE Software Editor in Chief for the inclusion of data science as a key knowledge area into “The Guide to Software Engineering Body of Knowledge” \cite{Ozkaya2020}. Modern methodologies such as Microsoft's Team Data Science Process (TDSP)\cite{TDSP2022} aim to integrate software and data processes and cover responsibilities for roles such as data scientists and application developers.} \par
\textcolor{black}{However, recent assessments by researchers assessed TDSP methodology as too focused on Microsoft technology and not sufficiently representative of the data scientist role\cite{martinez2021data}. The area of development, improvement and implementation of processes to support MDST teams and their projects offers much opportunity for further work.}
\subsubsection{\textcolor{black}{Tools for MDSTs}} 
Tools, workflows and data pipelines are being implemented to automate and increase the efficiency of data flowing from source systems to destination systems\cite{amershi2019software,9226314Bosch}. 
However, \textcolor{black}{our study} found that monitoring and visualisation tools require significant domain knowledge and configuration to be useful and do not address data quality challenges out of the box. This need for additional effort and \textcolor{black}{overall} experiences of our participants align with Altendeitering et al.'s 
assessment of 18 state of the art data quality tools. They found that most data quality tools in the market are focused on definition, measurement and improvement. Analysis features such as root cause analysis, and integration features which include collaboration, remediation/curation and orchestration remain underdeveloped\cite{altendeitering2022functional}.\par 
\textcolor{black}{There is emerging research about tools and techniques in the area of metadata management and data catalogues \cite{10.1145/3448247groger}.} Understanding data in data-intensive solution development requires a deep understanding of the target questions to be answered by the solution and also the source data domain and context of data collection. \textcolor{black}{One of the causes of MDSTs having to deal with data challenges as identified in our study is the \textit{lack of description of data}, and these tools could facilitate better collection and management of this information}. 
Other areas of emerging research include: domain knowledge elicitation\cite{10.1145/3397481.3450637park} and visual domain modelling tools\cite{khalajzadeh2021bidaml}. 
\textcolor{black}{However, caution, and empirical evaluation, is required to ensure these tools will be adopted.} Studies exploring collaboration tools used in inter-disciplinary teams \textcolor{black}{have identified that adoption of these tools is not easy. The studies} found that data scientists and domain experts \cite{mao2019data, 10.1145/3397481.3450637park} experience adoption challenges \textcolor{black}{and MDSTs} tend to settle on using tools that everybody is familiar with, rather than expend effort to use expert tools from their own discipline or use new and innovative tools that may be more effective\cite{mao2019data}.\par
Our study did not identify any effort being made in adopting new, innovative tools to manage data challenges as supported by research, potentially likely because these tools are still missing functionality to support collaboration, curation and orchestration as well as the inherent adoption challenges. 
\subsubsection{\textcolor{black}{Cross-skilling and boundary objects in MDSTs}} 
Research on software engineering team performance is still exploratory in nature and has largely focused on characteristics that increase the productivity of teams\cite{dutra2015characteristics, kanij2014performance}.
However, the challenges facing multi-disciplinary teams are not just about productivity. Multi-disciplinary project teams lack common background knowledge and team members are accustomed to different working practices. They need to develop a shared understanding and ways of working. The team usually starts to develop new 'knowledge boundaries' around the actions that need to be performed and thereby create new ways of working\cite{RATCHEVA2009206}.\par
Building shared knowledge within the software development context has been explored in a job rotation study between support and development team members. The study found that learning new skills was cognitively high demanding, and whilst the new knowledge allowed participants to understand larger aspects of organisational processes, the job rotation resulted in redundant knowledge that was not valued as highly as specialist knowledge\cite{FAEGRI20101118}. \textcolor{black}{Our study contradicts this finding.  Our study found that team members that have cross-skills were seen as highly valued by team managers and executives.}\par
\textcolor{black}{However, our study did identify challenges regarding how to achieve successful cross-skilling. Our  participants reported that their efforts to encourage and offer support to} cross-skill data scientist and software engineer team members are not yet effective, and efforts to cross-skill non-technical team members into technical roles are also challenging. Our participants also identified that career paths may need to be developed for people with cross-skills. \textcolor{black}{Prior studies by Kim et. al.\cite{kim2017data} identified that software engineers can and do cross-skill as data scientists on their own accord. Further empirical studies that focus on evaluating the impact that cross-skilled specialisations could have on MDSTS dealing with data challenges would be useful and could help develop career and training paths for specialised DI experts}.\par
\textcolor{black}{Alternative to cross-skilling, the related theories about using boundary objects to generate shared knowledge are also relevant to our study. Developing cross-boundary knowledge is} challenging because it requires team members to shift their focus from their specialisation and existing optimised practices and instead expend effort to develop special boundary knowledge objects\cite{carlile2002pragmatic}. In their case study, Carlile et. al developed a pragmatic approach to building  new cross-boundary knowledge which includes re-direction of effort to create repositories with shared definitions, and standardised methods and objects for individuals to specify and learn about dependencies across boundaries\cite{carlile2002pragmatic}. 
\textcolor{black}{The nature of} boundary objects has been investigated in the context of software development practices. Pareto et. al. identified requirements that a system architecture document should satisfy to serve as a boundary object in systems and design work\cite{10.5555/1929101.1929141pareto}. Guidelines for managing boundary objects including their properties, how they are used, and how they can be supported by a tool and managed as artefacts have been developed in a design science study for the context of automotive agile systems engineering\cite{https://doi.org/10.1002/smr.2166wohlrab}. 
Our study \textcolor{black}{found usage of} boundary object documentation. Participants discussed communication skills, \textcolor{black}{artefacts such as presentations, diagrams and whiteboard drawings} and the ability to explain concepts in terms that the other party could understand.
\subsubsection{\textcolor{black}{Personality and culture for MDSTs}}
Personality impacts have been extensively studied in software teams\cite{cruz2015forty}, including impact on requirements engineering\cite{askarinejadamiri2016personality}, coding and pair programming\cite{salleh2014investigating}, and testing\cite{kanij2013empirical}. The impact of personality on multi-disciplinary teams, especially those involved in DI solution development, is little explored to date\cite{bonesso2022emotional}.\par
The impact of organisational culture as a mediating factor of big data analytics in firm performance has also been studied empirically and findings indicate that there are limits to what can be achieved with culture of knowledge individually. 
 Instead, systematic approaches for treating knowledge and culture are required to drive performance\cite{UPADHYAY2020}. 
 \textcolor{black}{Our findings align with this study. Whilst culture was identified as being very important to dealing with data challenges by participants, and a number of participants expressed that they had a positive team culture, yet they still had to deal with data challenges i.e. a good culture by itself is not enough evidenced by the fact that MDSTs continue to spend much effort dealing with data challenges.} 
\subsection{Recommendations \textcolor{black}{for Software Engineering Researchers and Practitioners}}

\textcolor{black}{Software Engineering practitioners and researchers are well versed in the development and use of tools and processes and techniques to support software delivery. Hence, Software Engineers are in a unique position to use this expertise and lead the support of MDSTs. Our findings regarding Data-intensive software engineering and data-quality visualisation and monitoring show that this is already happening, but much more needs to be done to develop tools and associated processes. Our study offers the following recommendations:}
\vspace{0.2cm}\\
\textcolor{black}{\textbf{Recommendation 1: Align tools with methods and improve support for MDSTs in dealing with Data Challenges. }} Software engineering practitioners participants discussed tools they developed or were planning to implement, including data quality visualisation tools and data quality monitoring tools and methods. There were many industry standard data analysis and data quality tools available in the market \cite{ehrlinger2019survey}, with continuous ongoing development. \textcolor{black}{However, these tools are not yet domain specific and do not come with guidelines and processes, and so they stop short of supporting the actual dealing with data challenges. Features that facilitate domain specific knowledge capture, sharing and reuse are required to support data understanding. Features are also required to extend tools to support root cause analysis, and collaboration, remediation/curation and orchestration. Features to support data alignment from multiple sources and decisions regarding alignment are required.}
\\
\textbf{\textcolor{black}{Recommendation 2: Enhance and evaluate methods with tools. }}
\textcolor{black}{Agile delivery was a popular method identified by our participants, and hence it may offer a useful base to incorporate specific techniques to support MDSTs dealing with data challenges. For example, techniques to elicit, capture, and model domain knowledge could be developed and aligned with an incremental, agile way of working.  Visual domain modelling tools \cite{Khalajzadeh2020} need to be further explored and integrated into an overall method and process. Tools need to be dynamic and fit into an agile delivery approach so that data challenges can be dealt with incrementally.} 
\\
\textbf{\textcolor{black}{Recommendation 3:} Make the needs of different disciplines a non-negotiable requirement for empirical evaluation of tools, methods and frameworks.} Researchers should develop and evaluate frameworks, methods, and tools in an MDST context. Applying design thinking, human-centred design \cite{Grundy2020} and model-driven approaches to offer opportunities for designing, building, and evaluating solutions for MDSTs. \textcolor{black}{Tools need to be usable and cater to the needs of the different types of end users in MDSTs, including domain, data and software experts i.e. not just data experts or just software engineers. More empirical evaluations across different roles are needed. Evaluation of tools, methods and frameworks should focus on whether communication or creation of boundary objects is reduced or better supported through the use of the tool, what level of 'cross-skilling' is required to use a tool effectively.} 
\\
\textbf{\textcolor{black}{Recommendation 4:} Cater for personality style when educating and cross-skilling software engineers. }Software engineers could benefit from more cross-skilling in data science concepts and techniques and in target application domains. \textcolor{black}{Due to their more introverted personality profiles this may not come naturally}. We heard from several participants that ``personality" of team members had a significance on the performance and quality of their work and work within the multi-disciplinary DI solution development team.  However, the influence of personality and indeed other human aspects e.g. gender, age, background, education, expertise, communication skills, etc. are still unclear in this context. Further empirical studies of the impact of human aspects would be worthwhile, potentially feeding into guidelines for software engineer education and practices for work in multi-disciplinary DI solution development teams, similar to how they have been investigated for requirements engineering \cite{hidellaarachchi2022influence}.
%
\subsection{Future Work}
This study has identified that multi-disciplinary teams developing data-intensive software solutions experience and have to deal with significant data challenges. \textcolor{black}{We plan to assess the state of the art in software engineering processes, techniques and tools and data project processes for helpful tools and processes. We plan to develop guidelines and tools to support MDSTs and to articulate gaps in the state of the art.} We see opportunities for software engineers to improve on and develop new \textcolor{black}{processes and process aligned} tools that are specifically developed to support multi-disciplinary teams in resolving these challenges. Future research will also address the context of multi-disciplinary software teams in general so that software engineers can work effectively in a multi-disciplinary team environment. This could include research studies that assess cross-skilling in MDSTs and identification and assessment of boundary objects to support MDSTs.\par
\textcolor{black}{On a broader note, we also hope that this study motivates future STGT studies to expand and modify this theory, or to develop theories in different contexts and conditions to compare and contrast findings.}
\section{Evaluation, Threats, and Limitations}
\label{sec:threats}
\textcolor{black}{A first reaction by experts in the field  may be that this study does not offer any new expert knowledge because many of these issues have been experienced so often and are known.  However, by conducting this empirical study, and identifying and exploring the core challenges we offer a valuable model for communication and thinking about multi-disciplinary teams dealing with data challenges and pave the way for framing and evaluating future research.}
\textcolor{black}{\subsection{\textbf{Evaluating the method application}} 
Application of the STGT method is evaluated through the demonstration of evidence of} \textit{credibility} and \textit{rigour}\cite{Hoda2021}. We \textcolor{black}{provide evidence} of \textit{rigour} through providing coding and memoing examples \textcolor{black}{(see Figure \ref{fig:STGTcode})} and sanitised quotes throughout our findings to substantiate coding. \textcolor{black}{Evidence of \textit{credibility} of method application is demonstrated by providing evidence of method understanding and its effective application\cite{Hoda2021}.  We provide evidence of \textit{credibility} by} providing details on participant recruitment, interview process and providing details of how we applied the method and performed interleaved interviewing, coding and memoing \textcolor{black}{in Section \ref{sec:researchmethod}.} \textcolor{black}{The concepts and relationships emerged through our action of coding, constant comparison and memoing.} We outlined decisions that we made in applying the basic and advanced data collection and analysis, and the selection of \textcolor{black}{and} structured coding \textcolor{black}{into the 6Cs coding model}. 
\textcolor{black}{\subsection{\textbf{Evaluating the theory outcome}}}
\textcolor{black}{A first reaction by experienced readers to STGT outcomes may be that the theory outcome is ‘unsurprising'\cite{Hoda2021}.  Rather than being disappointing, this is actually interpreted as a sign of the theory being \textit{relevant}. \textit{Relevance} is a key criterion for evaluating preliminary STGT outcomes consisting of concepts and preliminary theories that are presented at conclusion of STGT Basic Stage Data Collection and Analysis\cite{Hoda2021}}.\par

\textcolor{black}{For a mature theory outcome which results after the application of the Advanced Stage of Theory Development, the STGT outcome} should demonstrate advanced criteria such as \textit{novelty, usefulness, parsimony and modifiability}\cite{Hoda2021}. Our paper presents a \textit{novel} \textcolor{black}{theory about MDSTs dealing with data challenges. There are very limited empirical studies available in this area as evidenced in our discussion about related works in Section 2. The theory has been developed empirically and structured using the 6Cs model. Whilst it covers complex socio-technical concepts, it can be communicated succinctly. It connects }different areas of research in software engineering, organisational theory, data governance and data quality in a dense and compact manner and thereby demonstrates \textit {parsimony}.
The resulting \textcolor{black}{theory} is \textit{useful} and \textit{parsimonious} as it \textcolor{black}{provides a compact overview and can facilitate communication, reasoning and hypothesis development to target future work. The theory has been derived empirically through interviews with participants. It applies in the specific context of productionised systems developed by industry practitioners working in a mix of commercial and government organisations delivering different types of DI solutions.}
\textcolor{black}{Our theory could be \textit{modified} to accommodate new findings, including additional depth and could be extended to different contexts. For example, adding data challenges or contingencies, and by further clarifying or extending the context and conditions of application.}
\textcolor{black}{\textbf{\subsection{Threats and limitations}}}
Even though the participants were assured that the interviews would be anonymised, there is a threat that they did not speak freely in the interview.  Further, the first author performed all the interviews and coding of the interviews. To mitigate these threats, the researchers purposefully chose the STGT method as it allows for, and expects, researchers to have some expertise in their area. The interview questions were reviewed by the first three and fifth authors and the first three researchers conducted a practice interview and reviewed the coding of the practice interview prior to finalising the interviews. The researchers also conducted reviews of codes and concepts throughout the study to validate the concepts and sought the feedback and advice of the fourth author, who was invited to join the team later to provide expert STGT guidance, especially toward theory development. All the authors were involved in the writing of the manuscript.



Concepts are identified and related through discussions with participants. By applying the STGT theoretical sampling and constant comparison approach and reflecting emerging findings back to subsequent interview participants to seek additional perspectives and properties we mitigated risks to concept validity. \textcolor{black}{Nevertheless, because we are not using statistical sampling, there remains the possibility that inclusion of a new participant would identify something previously unconsidered. By continuing with the interviews until no further new concepts were found after interview P19, we decided that we had reached a point of saturation. We continued the advanced stage theory development stage until we felt that the model 'presented a reasonable accurate statement of the matters studied' and would be useful for others studying a similar area. Further, the creation of a grounded theory should be seen as a process, not as a completed, verifiable product\cite{GlaserBarney2017Tdog}. The findings of our study are limited to the information gathered from 24 participants, a broad group of participants representing a variety of domains, organisation types and sizes, roles, seniority and educational backgrounds. However, the findings specific to individual domains and roles have not been investigated at great depth. As typical of qualitative studies, the findings are limited to the contexts studied, which in turn are limited by our access to participants. 
 \textcolor{black}{Whilst the recruitment of the study was advertised relatively widely on Twitter and LinkedIn and hence open to participants globally, the final participants were mainly from Australia and New Zealand (22/24), with one additional participant from China and one from Europe.} While recruiting for interviews is acknowledged to be challenging, it was particularly so during the pandemic. Future studies can explore individual aspects of our theory in greater depth and breadth.}

\section{Conclusion}
\label{sec:conclusion}
We conducted a socio-technical grounded theory (STGT) study where we interviewed 24 practitioners in multi-disciplinary, data intensive software teams (MDSTs) about their experience of developing and delivering data-intensive software solutions. \textcolor{black}{We focused on the core concern of dealing with data challenges, and structured our empirical findings into the 6Cs model.} We found that MDSTs' main concern is having to deal with a variety of data challenges. We identified the challenges of: Gaining Access to Data, Understanding Data, Aligning Data and Resolving Data Quality issues. We identified a number of related aspects, such as the \textit{context} in and \textit{condition} under which these challenges arise and their adverse \textit{consequences} such as Inability to Achieve Expected Outcomes, Solution Delivery Delays and Poor Reputation and Lack of Trust in the results. We also identified a number of \textit{contingencies} or strategies that \textcolor{black} {practitioner are applying} to address them. For example, we found that leaders are implementing longer term strategic initiatives including data governance programs, and team managers are seeking to improve team member skills through cross-skilling and improving communication and explanation skills. We found that software engineers in MDSTs are implementing data quality tools to monitor and visualise data quality. 
\textcolor{black}{These findings elevate the need to deal with data challenges as a primary concern and provide valuable insights into the needs of MDSTs. There is a need for future tools and methods to incorporate means for dealing with, minimising or preventing data challenges. Evaluation of tools and methods should also be extended to be more sensitive to existing work practices and needs of multi-disciplinary teams}.


%



\section*{Acknowledgment}
This study was supported by funding from the ARC Laureate Fellowship FL190100035. 


\ifCLASSOPTIONcaptionsoff
  \newpage
\fi



\bibliographystyle{IEEEtran}
\bibliography{TSEPaperRef}
%



%
%

%

\begin{IEEEbiography}[{\includegraphics[width=1in,height=1.25in,clip,keepaspectratio]{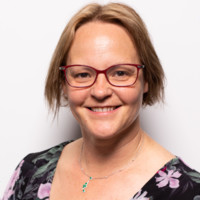}}]{Ulrike M.  Graetsch} is a PhD student at Monash University, Faculty of IT, Humanise Lab.  She completed her undergraduate Honours degree in Computing and Masters of Computing at Monash University.  Since graduating, she gained over 20 years of experience as a practitioner in IT Project Delivery and gained deep experience in implementing systems in different environments.  She has a keen interest in data analytics software delivery and ensuring that solutions meet the needs of diverse end-users.  The goal of her PhD is to improve capturing and modelling of human-centric requirements within multi-disciplinary teams that develop data analytics software.
\end{IEEEbiography}
\begin{IEEEbiography}[{\includegraphics[width=1in,height=1.25in,clip,keepaspectratio]{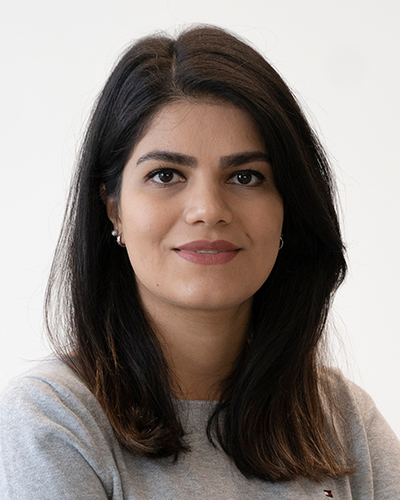}}]{Hourieh Khalajzadeh} is a Senior Lecturer in the School of Information Technology at Deakin University. Previously, she was a Research Fellow in the HumaniSE Lab at Monash University. Hourieh’s research is situated at the intersection of software engineering and data science. She is currently looking at the human-centric issues in Software Engineering and is experienced in designing domain specific visual languages for different applications, including big data analytics development. She has received several awards such as ACM Distinguished Paper Award (MobileSoft 2022), Outstanding Reviewer Award (CHASE 2021), Best Paper Award (ENASE 2020), and Best Showpiece Award (VL/HCC 2020). She serves as CHASE 2023 co-chair and ICSE 2023 Schedule Chair.
\end{IEEEbiography}

\begin{IEEEbiography}[{\includegraphics[width=1in,height=1.25in,clip,keepaspectratio]{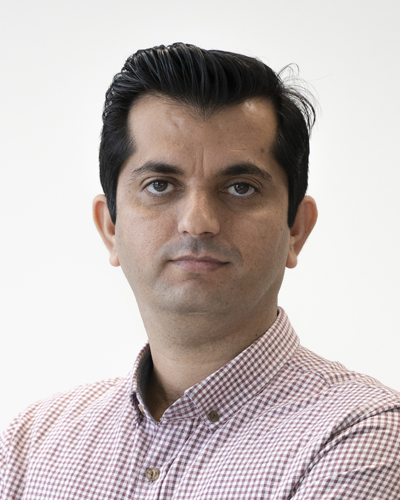}}]{Mojtaba Shahin} is a Lecturer in the School of Computing Technologies at RMIT University, Melbourne. Previously, he was a Research Fellow at Monash University. His research interests reside in Empirical Software Engineering, Human and Social Aspects of Software Engineering, and Secure Software Engineering. He completed his PhD study at the University of Adelaide, Australia.
\end{IEEEbiography}

\begin{IEEEbiography}[{\includegraphics[width=1in,height=1.25in,clip,keepaspectratio]{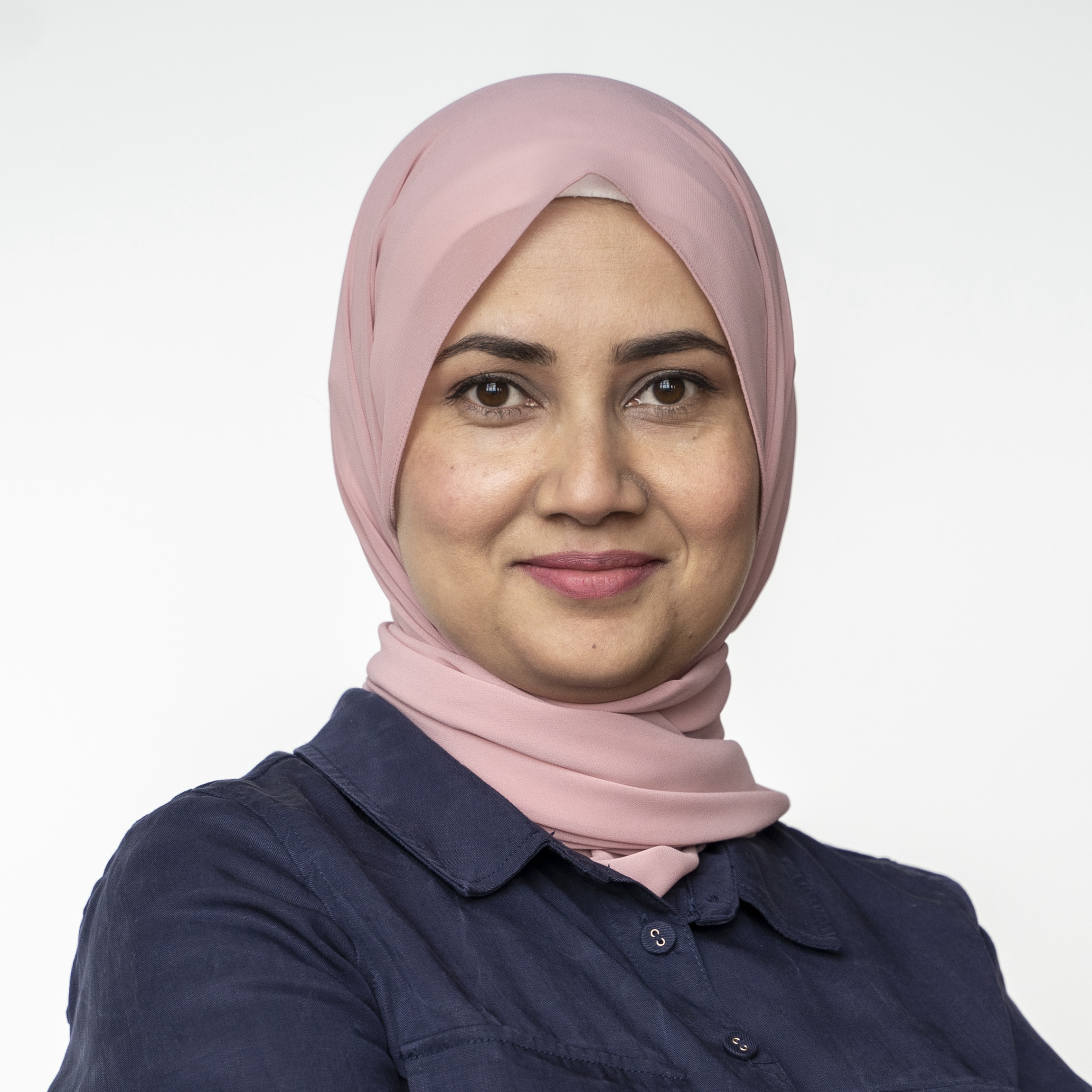}}]{Rashina Hoda} is an Associate Professor in Software Engineering at Monash University, Australia. Rashina specialises in human-centered empirical software engineering and has introduced socio-technical grounded theory (STGT) for software engineering. She received an ACM SIGSOFT Distinguished Paper Award (ICSE 2017) and Distinguished Reviewer Award (ICSE 2020). She serves as an Associate Editor of the IEEE Transactions on Software Engineering and the PC co--chair of the SEIS track of ICSE2023. Previously, she served on the IEEE Software Advisory Board, as Associate Editor of Journal of Systems and Software, CHASE 2021 PC co--chair and XP2020 PC co--chair. More details on https://rashina.com.
\end{IEEEbiography}

\begin{IEEEbiography}[{\includegraphics[width=1in,height=1.25in,clip,keepaspectratio]{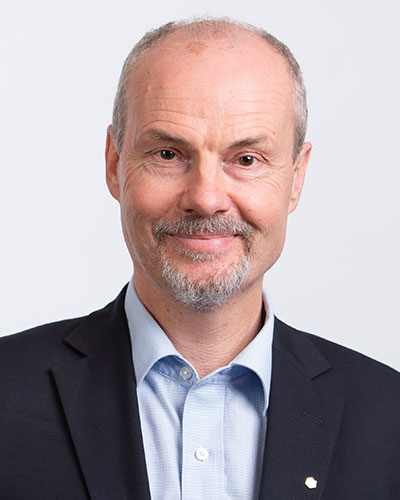}}]{John Grundy} is an Australian Laureate Fellow and a Professor of Software Engineering at Monash University where he heads up the HumaniSE lab in the Faculty of Information Technology.  His lab investigates “human-centric” issues in software engineering – these include, but are not limited to, impact of personality on software engineers and users; emotion-oriented requirements engineering and acceptance testing; impact of different languages, cultures and belief sets on using software and engineering software; usability and accessibility of software, particularly for ageing people and people with physical and mental challenges; issues of gender, age, socio-economic status and personal values on software, software requirements, and software engineering; and team and organisational impacts, including team climate. His lab's goal is to improve software engineering practices, tools, and thus the target systems to make better software for people.
\end{IEEEbiography}




\end{document}

%% file: tables/table1.tex
\begin{table*}
  \caption{Interview Participants Demographic Details}
  \label{tab:participants}
  \vspace{-2.5mm}
  \small
  \resizebox{17cm}{!}{
  \begin{tabular}{llllllll p{5cm}}
  \hline
 
\textbf{ID} & \textbf{Domain} & \textbf{Software Org} & \textbf{Country} & \textbf{Age} & \textbf{Gender} & \textbf{Highest Qualification}  & \textbf{Current Role} & \textbf{Prof. Exp.}\\
\hline \hline
P1 & Transportation & Yes & Australia   & 50  & Male   & Bachelor of Computer Science  & Data Architect & 25yrs software engineering \\
P2 & Parking Service & No & Australia   & 38  & Male   & Master of Education  & GM Data Analytics & 17yrs domain and data analytics\\
P3 & Health & No & Australia   & 45  & Male   & Graduate Diploma Business & Director of Data & 25yrs domain expertise \\
P4 & Health & No & Australia   & 30  & Female & PhD Computational Biology & Data Scientist & 2.5yrs data science\\
P5 & Consultancy & Yes & Netherlands & 43  & Female & Master of Construction  & Lead Data Scientist & 3yrs data science, 13yrs project management\\
P6 & Logistics& Yes & China & 35  & Male   & Master of IT  & Product Manager & 10yrs product management\\
P7 & Legal& Yes & Australia   & 60  & Male   & Master of Computer Science  & Managing Director & 31yrss domain expertise\\
P8 & Retail& No & Australia   & 25  & Male   & Master of Analytics & Data Analyst & 2yrs data analyst\\
P9 & Consultancy & Yes & Australia   & 41  & Male   & Bachelor of IT (Hons) & CEO  & 5yrs domain expertise, 15yrs software engineering\\
P10 & Finance & No & Australia   & 32  & Male   & Master of Business Intelligence  & Lead Data Analyst  & 7yrs data analytics\\
P11 & Mining & No & Australia & 34  & Female & PhD Computer Science   & Data Scientist  & 1.5yrs data science\\
P12 & Finance & No & Australia   & 39  & Male   & Bachelor of Computer Science & Technical Lead & 10yrs technical lead, 10yrs software engineering\\
P13 & Consulting & Yes & New Zealand & 37  & Male   & Bachelor of Engineering and Medicine & Development Manager & 5yrs software engineering\\
P14 & Health& No & New Zealand & 45  & Male   & PhD Computer Science  & CIO   & 15yrs data analytics\\
P15 & Accounting& Yes & New Zealand & 44  & Female & PhD Computer Science & Lead Engineer & 16yrs software engineering\\
P16 & Finance & Yes & Australia   & 46  & Male   & Bachelor of Software Engineering & Software Engineer & 20yrs software engineering\\
P17 & Health & No & Australia   & 66  & Male   & Medical Fellowship & Oncologist / Product Owner & 24yrs domain expertise\\
P18 & Supply Chain& Yes & Australia   & 42  & Female & Graduate Certificate - Supply Chain & Consultant & 20yrs domain expertise\\
P19 & Real Estate & Yes & Australia   & 27  & Male   & PhD Software Engineering & Data Developer & 4yrs software / data engineering\\
P20 & Finance & No & Australia & 38 & Female & PhD Computer Science & Lead Data Analyst & 5yrs data science\\
P21 & Construction & No & Australia & 44 & Male & Master in Business Administration & BI Manager &  20yrs data analysis\\
P22 & Consumer Goods & No & Australia & 50 & Male & Bachelor of Mathematics & Analytics Architect & 5yrs Analytics Architect, 16yrs data engineer\\
P23 & Consulting & No & Australia & 40 & Male & Masters of Business and Economics & Senior Manager & 16yrs domain expertise\\
P24 & Consulting & No & Australia & 33 & Male & Bachelor of Economics & Senior Manager & 11yrs data analytics\\
\hline
\end{tabular}}
\end{table*}